\documentclass[12pt]{article}

\usepackage{preprintstyle}
\usepackage[utf8]{inputenc}
\usepackage{comment}

\newcommand{\beq}{\begin{equation}}
\newcommand{\eeq}{\end{equation}}
\newcommand{\bal}{\begin{aligned}}
\newcommand{\eal}{\end{aligned}}
\newcommand{\bc}{\begin{cases}}
\newcommand{\ec}{\end{cases}}
\newcommand{\rmd}{\mathrm d}

\newcommand{\mc}{\mathcal}

\usepackage{dsfont}
\usepackage[normalem]{ulem}
\usepackage{cancel,xcolor}
\usepackage{makecell}
\setcellgapes{4pt}

\title{Limits on the Statistical Description of Charged de Sitter Black Holes}

\author{Lars Aalsma$^a$,}
\author{Puxin Lin$^b$,}
\author{Jan Pieter van der Schaar$^{c}$,}
\author{Gary Shiu$^b$,}
\author{Watse Sybesma$^{d}$}
\emailAdd{laalsma@d.umn.edu}
\emailAdd{plin73@wisc.edu}
\emailAdd{j.p.vanderschaar@uva.nl}
\emailAdd{shiu@physics.wisc.edu}
\emailAdd{watse.sybesma@su.se}
\affiliation{$^a$Department of Physics and Astronomy, University of Minnesota Duluth, 1049 University Drive, Duluth, MN 55812, USA}
\affiliation{$^b$Department of Physics, University of Wisconsin-Madison, 1150 University Avenue, Madison, WI 53706, USA}
\affiliation{$^c$Delta Institute for Theoretical Physics and Institute of Physics, University of Amsterdam, Science Park 904, PO Box 94485, 1090 GL Amsterdam, the Netherlands}
\affiliation{$^d$Nordita, KTH Royal Institute of Technology and Stockholm University, Hannes Alfvéns väg 12, 106 91 Stockholm, Sweden}

\abstract{The thermodynamics of de Sitter black holes is complicated by the presence of two horizons and the absence of a globally defined timelike Killing vector. The standard choice of the Gibbons-Hawking Killing vector is at odds with the interpretation of the surface gravity as an acceleration measured by a physical observer at rest. Focusing on four-dimensional Reissner-Nordström de Sitter black holes we show that this issue can be resolved by adopting a normalization originally proposed by Bousso and Hawking, which defines thermodynamic quantities relative to the unique freely-falling observer at a fixed radial coordinate. Within this framework, we derive new first laws for the black hole and cosmological horizon and re-examine the black hole's heat capacity. We find that the heat capacity remains finite in the near-extremal Nariai limit, thus averting a breakdown of the semi-classical thermodynamic description. However, the heat capacity does vanish in the cold limit, as expected, and for Nariai black holes in the ultracold limit, indicating that fundamental limitations on the statistical description persist in these regimes. We discuss the implications of our results for log-$T$ corrections to near-extremal de Sitter black holes.}

\begin{document}

\maketitle

\section{Introduction}
The study of black hole thermodynamics has provided deep insights into the nature of quantum gravity, suggesting a profound connection between spacetime geometry and the microscopic degrees of freedom that make up the black hole. The extension of these concepts to asymptotically de Sitter space introduces a rich and complex structure due to the presence of a cosmological horizon in addition to the black hole horizon(s). This results in several interesting features not present for black holes in asymptotically flat or anti-de Sitter space. For instance, the existence of both black hole and cosmological horizons leads to two notions of temperature. Due to the cosmological horizon, black holes in de Sitter space also have a maximum size.

A primary difficulty in defining the thermodynamics of de Sitter black holes stems from the lack of a globally defined timelike Killing vector. This limitation makes the definition of conserved quantities, such as the total mass or energy, subtle. From the perspective of an observer between the (outer) black hole horizon and the cosmological horizon, it is natural, however, to use a timelike Killing vector in the static patch that coincides locally to a static observer's local frame. The conventional approach by Gibbons and Hawking uses a normalization of this Killing vector given by $\xi = \partial_t$ \cite{Gibbons:1977mu}, where $t$ is the standard static time coordinate. In empty de Sitter space, this corresponds to normalizing the Killing vector to $-1$ at the pole of de Sitter space. 

However, as we will discuss, in the presence of a black hole this choice results in thermodynamic behavior seemingly at odds with the physical interpretation of the surface gravity of an horizon as an acceleration measured in the reference frame of a physically meaningful observer.\footnote{For black holes in asymptotically flat and anti-de Sitter space, it is conventional to normalize the Killing vector to $-1$ at spacelike infinity.} For instance, in the Nariai limit, where the (outer) black hole horizon and the cosmological horizon meet, the convention by Gibbons and Hawking leads to a vanishing temperature seemingly at odds with the finite temperature measured by an observer at the pole of the near-horizon dS$_2 \times S^2$ geometry. To remedy this, Bousso and Hawking adopted a normalization of the Killing vector that uses the perspective of the unique, freely-falling observer in between the (outer) black hole and cosmological horizon. At the radius $r=r_{\cal O}$, where the gravitational attraction of the black hole precisely cancels the cosmological repulsion, the Killing vector is normalized to $-1$ \cite{Bousso:1996au} (see also \cite{Morvan:2022ybp}). This normalization was later generalized to electrically charged Reissner-Nordström black holes \cite{Morvan:2022aon}. As we will show in detail, this choice of normalization reproduces the temperature of the near-horizon de Sitter geometry in the Nariai limit.

The Bousso-Hawking normalization requires a careful re-examination of the thermodynamical laws for black holes in de Sitter space\footnote{The thermodynamics of de Sitter black holes has been previously studied in \cite{Romans:1991nq,Mann:1995vb,Bousso:1996au,Booth:1998gf,Bousso:1999ms,Montero:2019ekk,Morvan:2022ybp,Morvan:2022aon,Castro:2022cuo}. Not all thermodynamic properties of black holes depend on the normalization of the Killing vector, but we'll see it is crucial when computing the heat capacity.}, as the total energy for example is no longer simply the mass parameter $M$ that appears in the metric in static coordinates \cite{Dolan:2013ft}. Instead, the physically relevant mass must be defined with respect to this new reference frame. In this paper, we focus on four-dimensional (electrically) charged Reissner-Nordström de Sitter black holes, although our results have a straightforward generalization to higher dimensions and black holes that rotate. We use the Bousso-Hawking normalization to analyze its thermodynamic properties, focusing on the Nariai, (ultra)cold and lukewarm limits. We review the black hole's characteristic `shark fin' phase diagram, which is bounded by two types of seemingly extremal solutions: cold black holes and charged Nariai black holes. These solutions intersect at the ultracold black hole. After that, we show how the thermodynamic laws are modified.

Our key contributions are twofold. First, we derive new first laws for the black hole and cosmological horizons using Bousso-Hawking normalization. In this derivation, we find it convenient to introduce an auxiliary York boundary at $r=r_{\cal O}$, which allows us to derive separate first laws for the black hole and cosmological horizon \cite{Banihashemi:2022htw}. In the final result, however, the dependence on the York boundary is absorbed into variations of the other thermodynamic quantities and the first laws relate variations of the observer-normalized mass $\tilde{M}$ to variations in the entropy and charge. Our definition of mass naturally incorporates the redshift factor between the traditional asymptotic mass $M$ and the observer's reference frame. Using this appropriately modified first law we classify the different thermodynamical regimes in the confined configuration space of charged de Sitter black holes, emphasizing the important role of the lukewarm solutions, corresponding to stable equilibrium states in the canonical ensemble (at fixed charge). 

Second, armed with the thermodynamic identities defined using Bousso-Hawking normalization we explore a potential breakdown of thermodynamics in near-extremal limits. In the classic paper \cite{Preskill:1991tb}, it was argued that the statistical description of near-extremal black holes breaks down when the (absolute value of the) heat capacity becomes smaller than one. At this point, the difference between the energy of the black hole (above the ground state) and the extremal energy drops below the typical energy of a Hawking mode, which suggests potentially large (quantum) corrections. This idea was indeed corroborated from the perspective of the Euclidean path integral, which revealed large quantum corrections to the partition function of near-extremal black holes \cite{Iliesiu:2020qvm,Heydeman:2020hhw,Iliesiu:2022onk,Banerjee:2023quv,Kolanowski:2024zrq,Kapec:2024zdj,Arnaudo:2024bbd}. These corrections find their origin in one-loop contributions around the saddle-point and modify the entropy to scale logarithmically with $T$, where $T$ is the temperature of the black hole, and are often referred to as log-$T$ corrections. They become large at low temperatures where they lead to a breakdown of the semi-classical approximation. We stress that in this paper we are not computing log-$T$ corrections. Instead, we use the heat capacity as an indicator where the semi-classical description breaks down.

With our refined thermodynamic approach, embracing the consequences of the Bousso-Hawking normalization, we reassess the behavior of the heat capacity of black holes in de Sitter space in the different extremal limits. We show that the heat capacity remains finite and large for a large part of the phase diagram of four-dimensional electrically charged Reissner-Nordström de Sitter black holes. In particular, we do not see a breakdown of the thermodynamic description of near-extremal Nariai black holes sufficiently far away from the ultracold point. This is to be contrasted with the behavior of the heat capacity computed using the normalization by Gibbons and Hawking, which goes to zero in the Nariai limit \cite{Castro:2022cuo}. We do find that the heat capacity goes to zero in the cold and ultracold Nariai limit, where the scaling of the heat capacity in fact reproduces the one obained using Gibbons-Hawking normalization.

The structure of this paper is as follows. In Sec. \ref{sec:thermo}, we introduce the Reissner-Nordström de Sitter metric and its phase diagram of regular solutions. We then detail the necessity and mechanics of the Bousso-Hawking normalization and derive the associated first laws.  Next, we compute the heat capacity in Sec. \ref{sec:HeatCap} and discuss the different extremal limits. We derive a list of the near-extremal behavior of the different thermodynamic quantities in Sec. \ref{sec:NearExt} and finally discuss the implications of our findings for the microscopic and statistical understanding of de Sitter black holes in Sec. \ref{sec:Conc}. Some mathematical details are put aside in the appendices.

{\bf Note added:} while we were completing this paper, we learned of the related work \cite{Chen:2025jqm} which studies the one-loop gravitational path integral of Reissner-Nordström de Sitter black holes. They focus on computing the phase of the partition function using a reduced two-dimensional dilaton gravity description, whereas we study the four-dimensional thermodynamics. Our results are complementary.

\section{Thermodynamics of Charged de Sitter Black Holes} \label{sec:thermo}
We first introduce the four-dimensional Reissner-Nördstrom de Sitter black holes. The Einstein-Maxwell action with positive cosmological constant $\Lambda_4 = 3/\ell_4^2$ is given by
\beq
I = \int\rmd^4x\sqrt{-g}\left[\frac1{16\pi G_4}\left(R-\frac6{\ell_4^2}\right) - \frac14 F_{ab}F^{ab}\right] ~,
\eeq
with $\ell_4$ the four-dimensional de Sitter radius. The equations of motion are given by
\beq
\bal
R_{ab} - \frac12g_{ab}R + \frac3{\ell_4^2}g_{ab} &= 8\pi G_4T_{ab} & ~,\\
\nabla_aF^{ab} &=0 ~.
\eal
\eeq
The stress tensor is given by
\beq
T_{ab} = F_{ac}F_b^{\,\,\,c} - \frac14g_{ab}F_{cd}F^{cd} ~.
\eeq
Focusing on electrically-charged asymptotically de Sitter black holes the relevant metric and gauge field are
\beq
\bal \label{eq:dS_BH_metric}
\rmd s^2 &= -f(r)\rmd t^2 + f(r)^{-1}\rmd r^2 + r^2\rmd \Omega_2^2 ~,\\
f(r)&= 1  - \frac{r^2}{\ell_4^2} -\frac{2G_4M}{r} + \frac{G_4Q^2}{4\pi r^2} ~, \\
A&= \left(-\frac{Q}{4\pi r}+\lambda\right)\rmd t ~.
\eal
\eeq
Here, $\lambda$ is a gauge parameter and the parameters $M$ and $Q$ are typically associated with the mass and charge of the black hole. However, as we will show this identification is somewhat subtle due to the choice of normalization of the timelike Killing vector.

\begin{figure}[t]
\centering
\includegraphics[scale=.9]{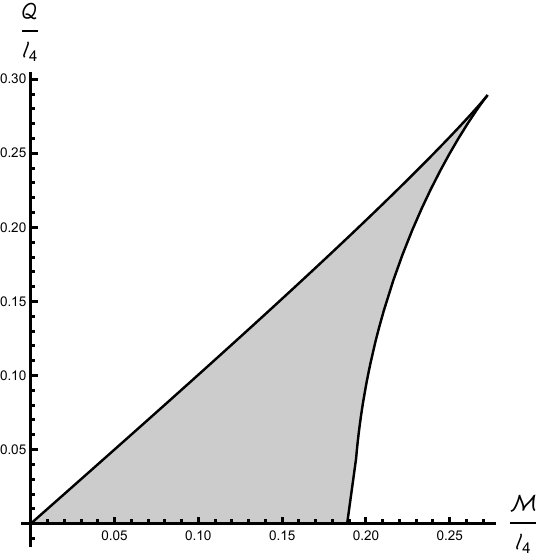}
\caption{Shark fin-shaped  diagram for Reissner-Nordström-de Sitter black holes. The shaded region corresponds to black hole solutions, and the white region to solutions with naked singularities.  The upper boundary of the shark fin corresponds to cold black holes $(r_a=r_b)$ and the lower curved boundary to Nariai black holes $(r_b=r_c)$. The two lines meet at the ultracold point $r_a=r_b=r_c$.}
\label{fig:sharkfin}
\end{figure}

As is well known, see e.g. \cite{Aalsma:2023mkz} and references therein, these black hole solution have a phase space of regular solutions that forms a compact `shark fin' (see Fig. \ref{fig:sharkfin}) bounded by three different notions of extremality. In terms of $(r_a,r_b,r_c)$, i.e. the inner black hole horizon, the outer black hole horizon and the cosmological horizon these extremal solutions are given by
\beq
\bal
\text{Cold Black Holes:} \quad r_a &= r_b  ~, \\
\text{Charged Nariai Black Holes:} \quad r_b &= r_c  ~, \\
\text{Ultracold Black Hole:} \quad r_a &= r_b =  r_c ~.
\eal
\eeq
Using the relation
\beq
\ell_4^2 = r_a^2+r_b^2+r_c^2 + r_ar_b+r_br_c+r_ar_c ~.
\eeq
the mass and charge parameters can be expressed in terms of two horizons and $\ell_4$ as
\beq
\bal
M &= \frac{{\cal M}}{G_4} = \frac{1}{2G_4}\frac{r_b^2(r_b^2-\ell_4^2)-r_c^2(r_c^2-\ell_4^2)}{(r_c-r_b)\ell_4^2} ~,\\
Q^2 &= \frac{4\pi {\cal Q}^2}{G_4} =  \frac{4\pi}{G_4}\frac{r_c(r_c^2-\ell_4^2)-r_b(r_b^2-\ell_4^2)}{(r_c^{-1}-r_b^{-1})\ell^2_4} ~,
\eal
\eeq
where we found it convenient to define the rescaled mass and charge $({\cal M},{\cal Q})$.

\subsection{Normalization of the Temperature}
To define the mass of the spacetime and other thermodynamic quantities we need to specify a timelike Killing vector. In global de Sitter space, this is subtle since no timelike Killing vector field exists. Also, there is no natural asymptotic boundary to define conserved charges. Instead, from the perspective of an observer in the same static patch region as the black hole it is natural to use the Killing vector $\xi \propto \partial_t$ that is timelike between the outer horizon of the black hole and the cosmological horizon. This still leaves the choice where to normalize the Killing vector to minus one, which we will discuss next.

For a general normalization we will write $\xi = \gamma\partial_t$. For asymptotically flat black holes, $\gamma$ is typically chosen to equal one, such that the norm of the Killing vector is normalized to one asymptotically. With this choice, the surface gravity at the black hole horizon has the interpretation of the acceleration needed to keep an object hovering at the horizon, as measured by a freely falling asymptotic observer at a fixed constant radius. Said differently, the locally defined (normalization-independent) acceleration
\beq
a^\mu = (\xi^\nu\nabla_\nu\xi^\mu)/(\xi^\rho\xi_\rho)  ~,
\eeq
which diverges at the horizon, leads to a finite surface gravity at the black hole horizon $r=r_h$ given by
\beq \label{eq:surfgrav}
\kappa = \lim_{r\to r_h} \sqrt{(a^\mu a_\mu)(\xi^\nu\xi_\nu)} ~,
\eeq
due to a vanishing redshift factor $\sqrt{-\xi^\mu\xi_\mu}$ between the horizon and asymptotic infinity.

Applying the same logic to de Sitter space leads to a wrinkle, as there is no analogous asymptotic region in de Sitter space. Nonetheless, Gibbons and Hawking \cite{Gibbons:1977mu} suggested to again pick $\gamma=1$. This is a natural choice for empty de Sitter space, as the location where $\xi^\mu\xi_\mu = -1$ corresponds to the pole of de Sitter space at $r=0$. In empty de Sitter space, due to the de Sitter isometry group, this is equivalent to the proper time of all free-falling observers. 

This gives the surface gravity of the (observer-dependent) de Sitter horizon an interpretation as the local acceleration times a redshift factor between the freely falling observer and the horizon. As we will show, this interpretation is no longer valid when including a black hole and still selecting $\gamma=1$.

The location where the norm of the Killing vector equals $-1$ is given by $\xi^\mu \xi_\mu = -\gamma^2f(r) = -1$.  Then, assuming $r\neq 0$ the quartic polynomial to solve to find this location is 
\beq \label{eq:quartic}
ar^4 + br^3 + cr^2 + dr + e = 0 ~,
\eeq
with
\beq
\bal
a &= 1~, \\
b &= 0 ~,\\
c &= \ell_4^2(\gamma^{-2}-1) ~, \\
d &=(r_a+r_b)(r_a+r_c)(r_b+r_c) ~, \\
e &= -r_ar_br_c(r_a+r_b+r_c) ~.
\eal
\eeq
The general solution to \eqref{eq:quartic} is given in App. \ref{app:quartic}. 

Now let us analyze the roots for different choices of normalization. If we take $\gamma = 1$, the determinant of this polynomial is negative for $r_c\geq r_b\geq r_a \geq 0$ such that there are only two real roots. For a general charged solution, we find that the positive root lies at a location $r<r_a$. In addition, this location corresponds to an accelerating trajectory with $a^\mu a_\mu \neq 0$. We therefore conclude that when choosing $\gamma=1$, the surface gravity of the black hole is defined with respect to an accelerating observer that lies behind the inner black hole horizon. Obviously, this suggests we use a different, more appropriate, normalization.

In the presence of a (charged) black hole, there is a unique location between the outer black hole and cosmological horizon where an observer is located at a fixed radius, free-falling and stationary with respect to the two horizons. This was pointed out for uncharged black holes in \cite{Bousso:1996au} and also applied to charged black holes in \cite{Morvan:2022aon}, where it was referred to as `static sphere' normalization (see also \cite{Faruk:2023uzs}). In this paper, we will refer to this choice as Bousso-Hawking normalization. One can think of this locus as the trajectory where the gravitational attraction of the black hole precisely cancels the cosmological expansion. This special stationary location, fixed between the two horizons, is unstable as free-falling observers at a slightly smaller radius fall through the black hole horizon and observers at a slightly larger radius are expelled through the cosmological horizon. The latter group of observers quickly see the black hole disappear (in a scrambling time) behind the cosmological horizon and settle down in empty de Sitter space. The other group of observers ends up at the black hole singularity. As we are interested in studying black hole states embedded in de Sitter space, it appears reasonable to focus on the unique observer that can probe both horizons. This special radius $r=r_{\cal O}$ is defined by $f'(r_{\cal O})=0$, which results in a quartic polynomial with one positive root that lies between the outer black hole horizon and the cosmological horizon. Normalizing the Killing vector to $-1$ at this location means that it coincides locally to the static observer's worldline, and fixes the normalization factor to $\gamma = 1/\sqrt{f(r_{\cal O})}$. Also note that this particular choice conveniently reduces to the standard normalization in both the empty de Sitter limit and the limit of vanishing cosmological constant ($\ell_4 \rightarrow \infty$). 

Writing the black hole temperature for an arbitrary normalization as $T_\gamma = \kappa_\gamma/(2\pi)$, we see that the relation between two choices of normalization $\gamma_1$ and $\gamma_2$ is given by
\beq\label{eq:tolmanlink}
T_{\gamma_2} = T_{\gamma_1}\left(\frac{\gamma_2}{\gamma_1}\right) ~.
\eeq
This is of course precisely the expression for the Tolman temperature \cite{Tolman:1930zza} that expresses a constant temperature $T_0$ in terms of the temperature seen by an (accelerating) observer that is redshifted by an amount $\gamma_1/\gamma_2$. This is the local temperature indicated by a thermometer held by this observer.

\subsection{Extremal and Near-Horizon Limits}
The choice of normalization is especially important when discussing the thermodynamics of black holes in de Sitter space. As we already alluded to, different choices of normalization lead to different values of the surface gravity at the black hole and cosmological horizon. For an arbitrary normalization, the temperature of the black hole and cosmological horizon are
\beq
T_{b,c} = \frac{\kappa_{b,c}}{2\pi} = \frac{\gamma}{4\pi \ell_4^2}(r_c-r_b)\left[2+\left(1+\frac{r_{c,b}}{r_{b,c}}\right)^2-\left(\frac{\ell_4}{r_{b,c}}\right)^2\right] ~.
\eeq
Using the normalization $\gamma = 1 $, this temperature vanishes in the Nariai limit $r_b\to r_c$ seemingly at odds with the finite temperature typically associated with the limiting dS$_2\times S^2$ spacetime. This reflects the fact that the standard normalization is defined with respect to an accelerating trajectory behind the black hole horizon, beyond the validity of the (rescaled) static coordinates. 

Indeed, these considerations suggest that the finite temperature of de Sitter space is measured by a non-accelerating observer that sits at the pole of the near-horizon dS$_2$ geometry. The Killing vector $\xi$ should therefore be normalized accordingly. To explicitly demonstrate this, let us introduce the following (proper) time and radial coordinates adapted to this special free-falling observer at the static sphere radius $r_{\cal O}$.
\beq
\bal \label{eq:NHcoordinates}
\tau &= \sqrt{f(r_{\cal O})} \,t ~, \\
\rho &= \frac{r-r_{\cal O}}{\sqrt{f(r_{\cal O})}} ~.
\eal
\eeq
Expressing the metric \eqref{eq:dS_BH_metric} in these coordinates we obtain
\beq
\rmd s^2 = -\left(\frac{f(\rho)}{f(r_{\cal O})}\right)\rmd\tau^2 + \left(\frac{f(\rho)}{f(r_{\cal O})}\right)^{-1}\rmd\rho^2 + r(\rho)^2\rmd\Omega_2^2 ~.
\eeq
In general, this metric will look rather complicated. However, let's expand around the Nariai limit. Since the static sphere location $r=r_{\cal O}$ is found by solving $f'(r) = 0$, recall also the discussion above \eqref{eq:tolmanlink}, we can expand $f'(r)$ as follows in the Nariai limit
\beq
\ell_4^2 r^3 f'(r) = f_0 + f_1(r_c-r_b) + f_2(r_c-r_b)^2 + {\cal O}(r_c-r_b)^3 ~,
\eeq
where the coefficients are given by
\beq
\bal
f_0 &= -2 \left(r-r_c\right) \left(-\ell_4^2 r_c+r^2 r_c+r r_c^2+3 r_c^3+r^3\right) ~, \\
f_1 &= \left(r-2 r_c\right) \left(\ell_4^2-6 r_c^2\right) ~, \\
f_2 &= -4 \left(r-2 r_c\right) r_c ~.
\eal
\eeq
This leads to the following near-horizon Nariai value for $r_{\cal O}$
\beq
\bal \label{eq:NearNariaiExp}
r_{\cal O} &= r_c - \frac12(r_c-r_b) + \frac{\ell_4^2-4r_c^2}{4r_c(6r_c^2-\ell_4^2)}(r_c-r_b)^2 + {\cal O}(r_c-r_b)^3 ~, \\
f(r_{\cal O}) &= \frac1{4r_c^2\ell_4^2}(r_c-r_b)^2(6r_c^2-\ell_4^2) + {\cal O}(r_c-r_b)^3 ~.
\eal
\eeq
Note that these expansions are not valid around the ultracold point where $r_b = r_c = \frac{\ell_4}{\sqrt{6}}$.

Plugging this expansion into the metric and taking the limit $(r_b,r_c)\to r_N$ we find
\beq
\rmd s^2 = -\left(1-\frac{\rho^2}{\ell_2^2}\right)\rmd\tau^2 + \left(1-\frac{\rho^2}{\ell_2^2}\right)^{-1}\rmd\rho^2 + r_N^2 \rmd\Omega_2^2 ~,
\eeq
where we defined the two-dimensional de Sitter radius
\beq \label{eq:dS2radius}
\ell_2 := \frac{\ell_4}{\sqrt{6-\ell_4^2/r_N^2}} ~.
\eeq
We recognize this as the metric of dS$_2\times S^2$ in static coordinates where the pole is located at $\rho =0 $, which corresponds to $r=r_N$. In the Nariai limit $r_{\cal O}\to r_N$ we see that the temperature defined using the normalization $\gamma=1/\sqrt{f(r_{\cal O})}$, given by
\beq
\tilde T_b = \frac{\sqrt{6r_c^2-\ell_4^2}}{2\pi r_c\ell_4} = \frac1{2\pi \ell_2}  , 
\eeq
indeed corresponds to the temperature measured by a freely falling observer at the pole of the near-horizon dS$_2$ space. Note that in this limit the isometries are enhanced implying that all free-falling observers in the dS$_2$ geometry are equivalent, whereas in the original geometry, outside the limit, the free-falling observer at $r=r_{\cal O}$ is special.  

On the cold branch, the temperature goes to zero independent of the choice of normalization, as long the normalization factor $\gamma$ does not diverge in this limit. This is clearly not an issue for the Bousso-Hawking normalization $\gamma = 1/\sqrt{f(r_{\cal O})}$.

Another limit of interest is the so-called lukewarm solution, where the black hole and cosmological temperatures coincide. This limit is independent of the choice of normalization and, in terms of the horizon radii, given by $r_b+r_c = \ell_4$. This translates to a charge-to-mass ratio of
\beq
\left(Q/M\right)_{r_b+r_c=\ell_4} = 2\sqrt{\pi G_4} ~,
\eeq
which, amusingly, is precisely the charge-to-mass ratio of a four-dimensional extremal Reissner-Nördstrom black hole in flat space\footnote{As such they can be interpreted as finite temperature generalizations of extremal black holes in flat space}. Lukewarm black holes have particularly nice properties, in particular the existence of regular Euclidean continuations and stability in the absence of charged decay. 

\begin{figure}[t]
\centering
\includegraphics[scale=.9]{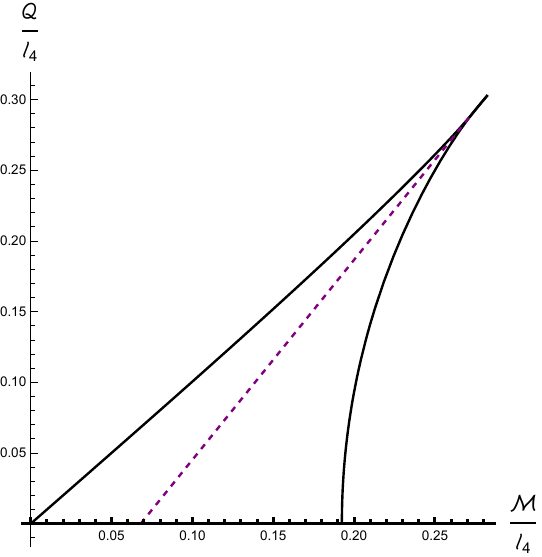}
\caption{There is a unique linear trajectory (purple, dashed) that remains within the sharkfin and ends at the ultracold point given by ${\cal Q} = \sqrt{2}{\cal M}-\frac{\ell_4}{6\sqrt{3}}$.}
\label{fig:UltraLinePlot}
\end{figure}

Finally, let us discuss the ultracold point. As is clear from \eqref{eq:dS2radius}, in this limit $\ell_2\to\infty$ such that the near-horizon geometry is Mink$_2\times S^2$. This limit is peculiar, foremost because there is an infinite scale separation between the Minkowski space and the two sphere at odds with general expectations \cite{Lust:2019zwm}. Furthermore, if we are studying the behavior of near-ultracold solutions it is important to ensure that the variations we consider do not exit the sharkfin. In fact, expanding around the ultracold solution, one finds that there is a unique linear trajectory that describes regular solutions that end up in the ultracold point, see Fig. \ref{fig:UltraLinePlot}.

In terms of mass and charge parameters this trajectory is given by
\beq
Q = 2\sqrt{2\pi G_4}M - \ell_4\sqrt{\frac\pi{27 G_4}} ~.
\eeq
Thus, when we are considering variations away from the ultracold point it is necessary to consider variations that obey $\partial Q/\partial M = 2\sqrt{2\pi G_4}$.\footnote{This was already observed in \cite{Morvan:2022aon,Castro:2022cuo}.}

\subsection{Derivation of First Laws with Bousso-Hawking Normalization}
It is now of interest to understand how the thermodynamic laws governing the black hole and cosmological horizon depend on the choice of normalization. The precise form of the first law depends on the region under consideration. In \cite{Dolan:2013ft} the authors derived the first law by considering a Hamiltonian formalism. By integrating conserved quantities from the black hole horizon to either the cosmological horizon or infinity, they obtained different identities.

For instance, integrating from the black hole horizon to the cosmological horizon \cite{Dolan:2013ft} found the following variational identity
\beq \label{eq:1stLawBetween}
T_b\delta S_b + T_c\delta S_c +\left(\Phi_b-\Phi_c\right)\delta Q = 0 ~.
\eeq
Here, the electric potential is defined as
\beq\label{eq:elpot}
\Phi_{b,c} = -\left.\xi^\mu A_\mu\right|_{r=r_{b,c}} = \gamma\left(\frac{Q}{4\pi r_{b,c}}-\lambda\right) ~,
\eeq
where we kept an explicit gauge parameter $\lambda$ and the corresponding entropies are given by the usual expression
\beq
S_{b,c} = \frac{\pi r_{b,c}^2}{G_4} ~.
\eeq
The electric charge is given by the integral
\beq
Q = \int_{S^2} \star F ~.
\eeq
Because the normalization of the Killing vector appears as an overall factor in \eqref{eq:1stLawBetween}, it drops out and this identity is valid for an arbitrary normalization.

By considering the region between either the black hole or cosmological horizon and infinity, \cite{Dolan:2013ft} also considered a first law in which the variation $\delta M$ appears, a variation of the mass parameter appearing in the metric \eqref{eq:dS_BH_metric}. The appearance of $\delta M$ implies the $\gamma = 1$ normalization and the first law is then given by \cite{Dolan:2013ft}
\beq
\bal \label{eq:GH_FirstLaw}
\delta M &= +T_b\delta S_b + (\Phi_b-\Phi_\infty)\delta Q ~, \\
\delta M &= -T_c\delta S_c + (\Phi_c-\Phi_\infty)\delta Q ~.
\eal
\eeq
We note that the derivation of this first law integrates over a region that goes outside of the (static) region in between the (outer) black hole and cosmological horizon. As such, it is not naturally adapted to the region accessible to a physical observer.
 
Upon integrating, it can be shown that this first law is compatible with a Smarr formula given by
\beq
\bal
M &=+2T_bS_b +(\Phi_b-\Phi_\infty)Q - 2P_\Lambda V_b ~, \\
M &=-2T_cS_c +(\Phi_c-\Phi_\infty)Q - 2P_\Lambda V_c ~,
\eal
\eeq
where $P_\Lambda = -\frac{\Lambda}{8\pi G_4}$ is a pressure term and $V_{b,c}=\frac{4\pi r_{b,c}^3}3$ is the thermodynamic volume. The derivation of the thermodynamic volume is given in App. \ref{app:thermoquantities}.

Using Bousso-Hawking normalization, one would expect similar first laws and Smarr formulae to hold. To derive the first laws, we find it convenient at an intermediate stage in the derivation to treat the trajectory of the free-falling observer at  $r=r_{\cal O}$ as a York boundary. We can then obtain a first law by following the approach of \cite{Dolan:2013ft} and integrate from the black hole horizon to the location $r = r_{\cal O}$. This implies that we are studying the thermodynamics of the region between the black hole horizon and a freely falling observer. One can study the region between the York boundary and the cosmological horizon in similar fashion by integrating from the York boundary to the cosmological horizon.

One important difference with the standard treatment of a York boundary, however, is the fact that variations of the location $r = r_{\cal O}$ depend on the black hole parameters. We can express (variations of) $r_{\cal O}$ in terms of (variations of) the mass and charge. For this reason, we will only use the formalism of York boundaries as a convenient computational tool but remove its explicit dependence in the final form of the first laws.

As was demonstrated in \cite{Banihashemi:2022htw} (see also \cite{Svesko:2022txo} for related two-dimensional work) using the Iyer-Wald formalism, an additional term appears in the first law when adding a York boundary. Writing quantities evaluated using Bousso-Hawking normalization with a tilde, one can obtain the following first laws for the Brown-York energy $E$
\beq
\bal \label{eq:BH-firstlaws}
\delta E &= +\tilde T_b\delta S_b +(\tilde \Phi_b-\tilde\Phi_{\cal O})\delta Q - P_{\cal O}\delta A_{\cal O} ~, \\
\delta E &= -\tilde T_c\delta S_c +(\tilde \Phi_c-\tilde\Phi_{\cal O})\delta Q - P_{\cal O}\delta A_{\cal O} ~.
\eal
\eeq
Here $P_{\cal O}$ is a pressure term associated with the York boundary, for the derivation and explicit form see App. \ref{app:thermoquantities}, and $A_{\cal O} = 4\pi r_{\cal O}^2$ is its area. The tildes indicate that these quantities depend explicitly on the normalization. The pressure and area term depend on the location $r_{\cal O}$, which is indicated by the the subscript.

At this point, a clarifying remark is appropriate. At first sight, the identification of $\delta E$ as the variation of the Brown-York energy does not seem to be straightforward. In \cite{Banihashemi:2022htw}, the first law involving the Brown-York energy and the York boundary was defined in terms of the normalization-independent Tolman temperature, which can be defined at any fixed radial location. Instead, the first law \eqref{eq:BH-firstlaws} was written down in terms of the Bousso-Hawking temperature. Therefore, the identification of $E$ as the Brown-York energy can only be made when putting the York boundary at $r=r_{\cal O}$, at which point these two (generally) different definitions of temperature agree. We explain this more in detail in Appendix \ref{app:BHenergy}.

By subtracting the first laws in \eqref{eq:BH-firstlaws} one can reproduce the first law for the region between the black hole and cosmological horizon \eqref{eq:1stLawBetween}, which is independent of the normalization. One can also obtain the following Smarr formulae \cite{Banihashemi:2022htw}
\beq
\bal
E &= +2\tilde T_b S_b + (\tilde\Phi_b-\tilde\Phi_{\cal O})Q - 2P_\Lambda\left(\tilde V_b-\tilde V_{\cal O}\right) -2P_{\cal O}A_{\cal O} ~, \\
E &= -2\tilde T_c S_c + (\tilde\Phi_c-\tilde\Phi_{\cal O})Q - 2P_\Lambda\left(\tilde V_c-\tilde V_{\cal O}\right) -2P_{\cal O}A_{\cal O} ~.
\eal
\eeq
Here $\tilde V_{r_{b,c,\mathcal{O}}} = \frac{4\pi r_{b,c,\mathcal{O}}^3}{3\sqrt{f(r_{\cal O})}}$ is the thermodynamic volume with Bousso-Hawking normalization.

As already mentioned, variations of the York boundary in our case are not independent variations and by comparing \eqref{eq:BH-firstlaws} with \eqref{eq:GH_FirstLaw} we see that (using $\Phi_\infty = 0$ by a suitable choice of gauge)
\beq \label{eq:YorkTildeRela}
\delta E + P_{\cal O}\delta A_{\cal O} + \tilde\Phi_{\cal O}\delta Q = \frac{\delta M}{\sqrt{f(r_{\cal O})}} ~.
\eeq
To absorb all explicit dependence on the York boundary, we therefore introduce a new energy variation $\delta \tilde M$ which we take to be
\beq \label{eq:NewMassVariation}
\delta \tilde M := \delta M/\sqrt{f(r_{\cal O})} ~.
\eeq
At this point this definition seems rather ad-hoc, but we will elaborate more upon the interpretation of this mass parameter in Appendix \ref{app:BHenergy}. There, we will argue that it is closely related to the Brown-York energy.

We have now removed all explicit dependence on the York boundary which yields our final result for the first laws
\beq
\bal \label{eq:FinalFirstLaw}
\delta \tilde M &= +\tilde T_b\delta S_b + \tilde \Phi_b \delta Q ~, \\
\delta \tilde M &= -\tilde T_c\delta S_c + \tilde \Phi_c\delta Q  ~.
\eal
\eeq

We are now interested in computing the black hole heat capacity in the different extremal limits. Before we do so, however, let us comment on the behavior of the electric potential.

\subsubsection{Finiteness of the Electric Potential}
Let us take a closer look at the behavior of the electric potential \eqref{eq:elpot} when we apply Bousso-Hawking normalization
\begin{equation}
	\tilde{\Phi}_{b,c}
    =
    \frac{1}{\sqrt{f(r_{\mathcal{O}})}}
	\left(
        \frac{Q}{4\pi r_{b,c}}
		-
		\lambda
	\right)
    =
	\frac{Q}{4\pi \sqrt{f(r_{\mathcal{O}})}}
	\left(
		\frac{1}{r_{b,c}}
		-
		\tilde{\lambda}
	\right)
	\,.
\end{equation}
In \cite{Castro:2022cuo} it was pointed out that since $\sqrt{f(r_{\mathcal{O}})}\sim (r_{b}-r_{c})+\mathcal{O}(r_{b}-r_{c})^{2}$ (see \eqref{eq:NearNariaiExp}) for $|r_c-r_b|\ll1$, the above expression diverges in the Nariai limit for the choice of $\tilde{\lambda}=0$.

Despite the electric potential not being directly observable, this could potentially signal a pathology. As we will show here, however, away from the ultracold point this diverging behavior is in fact inherited from subtracting the value of the electric potential at $r\to\infty$ instead of $r=r_{\mathcal{O}}$, translating to $\tilde{\lambda}=0$ or $\tilde{\lambda}=r_{\mathcal{O}}^{-1}$, respectively. Adopting $\tilde{\lambda}=r_{\mathcal{O}}^{-1}$ and using the expressions from \eqref{eq:NearNariaiExp}, we arrive at the expressions
\beq
\bal
\tilde{\Phi}_{b}
	&=+
	\frac{
		\ell_{4}Q
	}{
		4\pi r_{b}\sqrt{6r_{b}^{2}-\ell_{4}^{2}}
	}
    + {\cal O}(r_c-r_b) ~, \\
\tilde \Phi_c &=- 
\frac{
		\ell_{4}Q
	}{
		4\pi r_{c}\sqrt{6r_{c}^{2}-\ell_{4}^{2}}
	}
	+ {\cal O}(r_c-r_b) ~.
\eal
\eeq
We observe that the electric potentials are now finite in the Nariai limit. This is directly related to the finiteness of $\tilde{\Phi}_{b,c}-\tilde{\Phi}_{\mathcal{O}}$ that appeared in the previous section, as $\tilde{\Phi}_{b,c}-\tilde{\Phi}_{\mathcal{O}} = \left.\tilde \Phi_{b,c}\right|_{\tilde \lambda = r_{\cal O}^{-1}}$. In fact, in the Nariai limit we can nicely express the electric potential as
\beq
\lim_{(r_b,r_c)\to r_N}\tilde \Phi_{b,c} = \pm\left(\frac{\ell_2}{r_N}\right)\frac{Q}{4\pi r_N} ~,
\eeq
where $r_N$ is the Nariai radius and $\ell_2$ is the two-dimensional radius of the near-horizon de Sitter space. We recognize this as the electric potential for $\gamma=1$ normalization in the Nariai limit times the dimensionless factor $\ell_2/r_N$ stemming from the redshift factor.

In the ultracold limit $\ell_2/r_N \to \infty$ and the electric potential diverges, but, as we now explain, from the perspective of the two-dimensional near-horizon de Sitter space this is not surprising. The proper distance between the observer $r=r_{\cal O}$ at the pole of de Sitter space and the black hole horizon is finite and proportional to $\ell_2$. In the ultracold limit this proper distance diverges and it is more appropriate to express the electric potential with respect to a different reference point. 

Indeed, if we consider the gauge-independent field strength $F=\rmd A$ expressed in the natural near-horizon coordinates \eqref{eq:NHcoordinates} in the Nariai limit instead of the electric potential, we find that the electric field is independent of the normalization, finite and constant, as expected
\beq
F = -\frac{Q}{4\pi r_N^2} \rmd t\wedge\rmd r = -\frac{Q}{4\pi r_N^2} \rmd \tau\wedge\rmd \rho ~.
\eeq
We conclude that there are no pathological divergences in physical quantities using Bousso-Hawking normalization.

\section{Heat Capacity and Breakdown of Thermodynamics} \label{sec:HeatCap}
We have now understood how the thermodynamics of black holes in de Sitter space can be described for different choices of normalization of the timelike Killing vector. In particular, to properly define the thermodynamics using Bousso-Hawking normalization it seems useful to go through an intermediate step and introduce a York boundary at the location of the stationary free-falling observer between the black hole and cosmological horizon. After this, the relevant first laws can be written down in terms of redshifted quantities.

Doing so is particularly natural and useful in the Nariai limit, where the location of the observer at $r=r_{\cal O}$ corresponds to the origin (or pode) of the static patch of the emerging dS$_2\times S^2$ geometry. We will now be interested in using these results to study the (thermodynamic) stability of these (charged) black holes. One is immediately faced with a complication, however, due to the presence of two horizons that generically have different temperatures, prohibiting a description in terms of equilibrium thermodynamics. This difficulty was studied in some detail in \cite{Dinsmore:2019elr} for uncharged black holes and \cite{Johnson:2019ayc} also considered adding charge or rotation. 

The picture emerging from these works is that one can think of black holes in de Sitter space as constrained thermal subsystems in a de Sitter heat bath at a temperature given by the cosmological horizon. A similar interpretation was put forward in \cite{Morvan:2022ybp,Morvan:2022aon} where it was shown that the probability to nucleate a black hole from a de Sitter heat bath is given by the exponent of minus the total entropy of the system (the sum of the entropy of the black hole and cosmological horizon). The energy needed to nucleate the black hole comes at the expense of the de Sitter heat bath and such a (rare) fluctuation decreases the total entropy of the system.

Introducing a York boundary at the location of the observer at $r=r_{\cal O}$, essentially allowed us to treat the region between the black hole horizon and $r_{\cal O}$ and the region between $r_{\cal O}$ and the cosmological horizon as two separate thermodynamic systems. In fact, \cite{Banihashemi:2022jys} used a York boundary to rigorously be able to define the thermodynamic ensemble in de Sitter space and then taking the limit where the York boundary was removed. In our case, the York boundary arises naturally at the $r=r_{\cal O}$ and its contribution is absorbed into the other thermodynamic quantities.

It is of particular interest to study the limitations of this thermodynamic description. In a classic paper by Preskill et al. \cite{Preskill:1991tb} it was argued that the thermodynamic description of asymptotically flat Reissner-Nördstrom black holes breaks down close to extremality. At low temperatures, the thermal energy drops below the typical energy of a Hawking quantum leading to a situation analogous to a quantum freezeout in statistical systems. The inability to absorb heat due to absence of available degrees of freedom is indicated by a heat capacity at fixed charge that goes to zero in the extremal limit. The point where the statistical description becomes unreliable corresponds to a heat capacity of order one. More recent works (see \cite{Turiaci:2023wrh} for a review) have indeed confirmed that the near-extremal statistical description of Reissner-Nördstrom black holes is significantly modified due to large fluctuations around the saddle point in the Euclidean path integral. Taking these fluctuations into account leads to corrections to the entropy (and more generally the partition function) that scale logarithmically with temperature \cite{Iliesiu:2020qvm,Heydeman:2020hhw,Iliesiu:2022onk,Banerjee:2023quv,Kolanowski:2024zrq,Kapec:2024zdj,Arnaudo:2024bbd}.

Before we discuss the charged case, let us first consider the behavior of uncharged near-Nariai black holes. Using the first law for uncharged black holes, given by $\delta \tilde M = \tilde T_b \delta S_b$, we can see that the near-Nariai behavior of the mass is given by
\beq \label{eq:NearNariaiMass}
Q=0: \qquad \delta \tilde M = -\frac{\pi\ell_4^2}{G_4}\delta\tilde T_b + {\cal O}(\tilde T_b-\tilde T_N)^2 ~,
\eeq
where $\tilde T_N = (2\pi\ell_2)^{-1}$ is the temperature of the black hole in the Nariai limit. The analogous quantity for asymptotically flat Reissner-Nordström black holes is the (change in the) energy above the (extremal) ground state energy. In that case, the change in the energy above extremality can be interpreted as the energy available to produce Hawking modes. In the de Sitter case, we give \eqref{eq:NearNariaiMass}  the same interpretation as the energy available to Hawking modes, with the important difference that now there is a maximum energy state, which explains the minus sign.

This near-extremal behavior should be contrasted with the results of \cite{Castro:2022cuo} for de Sitter black holes, which were obtained using $\gamma = 1$ normalization. First, the dependence of the mass on temperature is linear instead of quadratic. Second, the available energy remains finite and large in the Nariai limit. Still assuming that the energy of Hawking modes have a linear dependence on the temperature, this shows that available energy never dips below the thermal energy. Hence, there is no indication that the thermodynamic description breaks down in the (uncharged) Nariai limit.

This conclusion can also be reached by examining the heat capacity of the black hole system. Using the first law, this heat capacity can be expressed as
\beq \label{eq:HeatCap}
\tilde C_b = \tilde T_b\left(\frac{\partial S_b}{\partial \tilde T_b}\right) = \frac{\partial \tilde M_b}{\partial \tilde T_b} ~.
\eeq
Using \eqref{eq:NearNariaiMass} we see that the heat capacity in the (uncharged) Nariai limit does not go to zero, but is large and given by
\beq \label{eq:UnchargedHeatNariai}
\tilde C_b = -\frac{\pi \ell_4^2}{G_4} ~,
\eeq
where the minus sign simply reflects the fact that the available energy is an energy \emph{below} the maximum Nariai mass.

The situation will be different when we include charge, as the temperature $\tilde T_b$ of Nariai black holes becomes smaller when increasing the charge. To assess the energy below the Nariai mass it is then natural to work at fixed charge.\footnote{As already mentioned, one needs to be careful near the ultracold point where variations at fixed charge exit the sharkfin.}  We therefore now turn our attention to computing heat capacities for the different extremal limits.

\subsection{Heat Capacity at Fixed Charge}
\begin{figure}[t]
\centering
\includegraphics[scale=1]{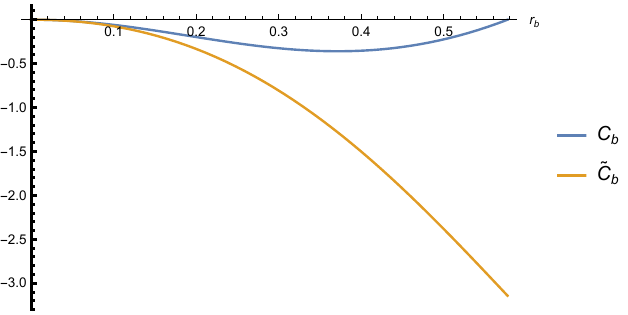}
\caption{Heat capacity of an uncharged black hole in de Sitter space for $\gamma=1$ normalization and Bousso-Hawking normalization as a function of radius. We used $\ell_4=G_4=1$.}
\label{fig:SH_BH_uncharged}
\end{figure}

Using the $\gamma=1$ normalization, it is easily demonstrated that the heat capacity of Reissner-Nördstrom de Sitter black holes also vanishes in the Nariai limit, see e.g. \cite{Castro:2022cuo,Blacker:2025zca}. One should not jump to the conclusion, however, that Nariai black holes have a similar breakdown in their statistical description. Using Bousso-Hawking normalization, the heat capacity is given by
\beq
\tilde C_b^Q = \tilde T_b\left(\frac{\partial S_b}{\partial \tilde T_b}\right)_Q ~,
\eeq
where $Q$ denotes that the charge is kept fixed. In Fig. \ref{fig:SH_BH_uncharged} we display the heat capacity as a function of the black hole radius. For comparison, we also include the heat capacity defined using $\gamma=1$ normalization

From Fig. \ref{fig:SH_BH_uncharged} we see that the heat capacity $\tilde C_b$ of an uncharged black hole is monotonically decreasing as a function of black hole radius. On one end, the heat capacity goes to zero for small (Planck mass) black holes. Concretely, in the semi-classical limit $\sqrt{G_4}/\ell_4 \ll 1$ we find that $\tilde C_b = -1$ for Planck mass black holes with radius $r_b \sim \sqrt{G_4}$. For these microscopic black holes, the energy of their thermal radiation becomes comparable to the energy of the black hole itself, which means that the system cannot be reliably described as ensemble at fixed temperature. This is also the point where general relativity as an effective field theory breaks down and higher curvature terms become relevant.

On the other end along the neutral branch, the Nariai limit   $r_b\to r_N = \frac{\ell_4}{\sqrt{3}}$, the heat capacity reduces to
\beq
\text{Uncharged Nariai:} \qquad \lim_{r_b\to r_N}\tilde C_b = - \frac{\pi \ell_4^2}{G_4} ~,
\eeq
which agrees with \eqref{eq:UnchargedHeatNariai}. The non-vanishing heat capacity implies that neutral Nariai black holes have a macroscopic number of states that could be excited, justifying the use of a semi-classical approximation. We conclude that, expressed in terms of the redshifted energy $\tilde M$, in the Nariai limit there are still a large number of degrees of freedom available that can absorb heat and the statistical description does not break down.

This picture changes when we include charge. By increasing the charge, we can lower the temperature of the black hole and we can end up in a situation where the available energy drops below the Hawking temperature. This is demonstrated very clearly for cold black holes which have a vanishing temperature along the entire cold branch. The near-extremal heat capacities are identical in the different normalizations. This can be attributed to the fact that the leading order term in an expansion of $f({r_{\cal O}})$ around the cold point is constant, as we will explain in more detail in Sec. \ref{sec:NearExt}. We find that the heat capacity at fixed charge is
\beq
\text{Cold:} \qquad \tilde C_b^Q = C_b^Q = \frac{\pi r_b^2}{G_4}\left(\frac{r_b}{r_a}-1\right) + {\cal O}(r_b-r_a)^2 ~.
\eeq
The vanishing of the heat capacity in the extremal limit suggests a breakdown of the thermodynamic description, similar to extremal black holes in flat space.

Note, however, that in flat space charged black holes thermodynamically evolve to the extremal state by emitting neutral Hawking radiation. In de Sitter space, cold de Sitter black holes are not final state of neutral Hawking radiation. At fixed charge, black holes evolve towards the maximal entropy lukewarm state instead. This is because, for masses below the lukewarm point, the cosmological temperature is higher than the black hole temperature, causing the black hole to evolve towards the lukewarm line instead of the cold extremal state. In other words, in de Sitter space one would need to do something special (throw in some extra charge or force the black hole to lose some mass) to end up in the non-equilibrium cold extremal state. 

\begin{figure}[t]
\centering
\includegraphics[scale=1.2]{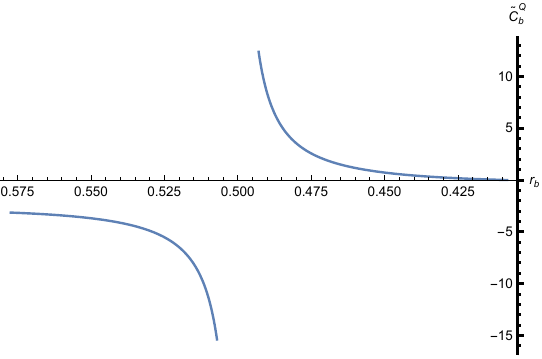}
\caption{Heat capacity at fixed charge as a function of radius of charged Nariai black holes in de Sitter space. We used $\ell_4=G_4=1$.}
\label{fig:SH_BH_charged}
\end{figure}

Turning to near-Nariai black holes, the heat capacity at fixed charge is given by
\beq
\bal
\text{Charged Nariai:} \qquad\tilde C_b^Q &= -\frac{\pi r_b^2}{G_4}\left(\frac{6r_b^2-\ell_4^2}{4r_b^2-\ell_4^2}\right) + {\cal O}(r_c-r_b) ~, \\
C_b^Q &= - \frac{\pi r_b^2}{G_4}\left(\frac{r_c}{r_b}-1\right) + {\cal O}(r_c-r_b)^2 ~.
\eal
\eeq
This expression reveals some interesting physics. First, we notice that at $r_b = \ell_4/2$, the heat capacity $\tilde C_b^Q$ diverges. This point corresponds to the intersection of the lukewarm line and the Nariai line at ${\cal Q}/\ell_4 = 1/4$. A divergent heat capacity is indicative of a phase transition \cite{Davies:1977bgr} (cf. \cite{Hut1977}). As will be discussed later, this intersection is associated with a continuous phase transition. Indeed, the qualitative behavior of the black hole states change when crossing the lukewarm line. Black holes above the lukewarm line (in Fig. \ref{fig:sharkfin}) have a temperature that is lower than that of the cosmological horizon. Below the lukewarm line, the heat capacity is negative and above the lukewarm line the heat capacity is positive, see Fig. \ref{fig:SH_BH_charged}\footnote{More generally one can show that at fixed charge below the lukewarm-Nariai intersection, a maximum black hole temperature point exists in between the lukewarm and the Nariai state. This is sometimes referred to as the Davies point \cite{Davies:1989}. Consequently, this corresponds to a transition from positive to negative heat capacity. These points form a line that intersects with the lukewarm and Nariai line exactly at the lukewarm-Nariai intersection.}. We will come back to the interpretation of this divergence in Sec. \ref{sec:LukewarmCap} and show how working in a different ensemble gives a finite and positive heat capacity.

Second, above the lukewarm line the heat capacity is positive and vanishes in the ultracold point $r_b=\ell_4/\sqrt{6}$. This suggests that the statistical description breaks down along the Nariai line when $\tilde C_b^Q < 1$, which corresponds to
\beq
r_b < \frac{\ell_4}{\sqrt{6}}\left(1+\frac{G_4}{\pi \ell_4^2}\right) + {\cal O}\left(G_4^2/\ell_4^4\right) ~,
\eeq
i.e. for black holes very close to the ultracold point. Approaching the ultracold point from the Nariai line this translates to a temperature of
\beq
\tilde T_b > \frac{\sqrt{3G_4}}{\ell_4^2\pi^{3/2}} + {\cal O}\left(G_4^2/\ell_4^4\right) ~.
\eeq
For temperatures smaller than this, we expect the thermodynamic description to break down. As the dS$_2$ section becomes infinitely large in the ultracold limit, this is probably connected to a strong coupling divergence in the IR of the reduced two-dimensional dilaton gravity description \cite{Chen:2025jqm}. 

One should be careful, however, when approaching the ultracold point. As explained, for black hole variations away from the ultracold point that remain within the sharkfin there is a unique trajectory. This implies that, when expanding around the ultracold point, one cannot consider variations where the charge is kept fixed. Instead, any variation should obey $\delta Q = 2\sqrt{2\pi G_4}\delta M$ corresponding to a non-standard thermodynamic ensemble. We will come back to this point, but for now we summarize the different heat capacities at fixed charge in Tab. \ref{tab:HeatCapCharge}.  

\begin{table}[t]
\centering
\makegapedcells
\begin{tabular}{| r || c c ||} 
 \hline 
 & $C_b^Q$ & $\tilde C^Q_b$ \\
 \hline\hline
Nariai:& $0$ & $-\frac{\pi r_b^2}{G_4}\left(\frac{6r_b^2-\ell_4^2}{4r_b^2-\ell_4^2}\right)$ \\
 \hline
Cold: & $0$ & $0$ \\
 \hline
Ultracold: & $0$  & $0$  \\
 \hline
Nariai $\cap$ Lukewarm: & $0$  & $\pm\infty$  \\
 \hline
\end{tabular}
\caption{Heat capacity at fixed charge of the different extremal black holes for both $\gamma=1$ and Bousso-Hawking normalization. The sign of the divergence in $\tilde C_b^Q$ at the intersection of the Nariai and lukewarm line depends if we approach that point from above or below.}
\label{tab:HeatCapCharge}
\end{table}

\subsection{Emission of Charged Particles} \label{sec:LukewarmCap}
The choice of ensemble for a statistical system determines what type of interactions it can have with the thermal bath, for instance whether energy, particle number or charge can be exchanged. Up to now, we only studied the heat capacity of black holes at fixed charge. This is appropriate for black holes undergoing Hawking radiation, which transfers only energy and heat. However, in addition to Hawking radiation charged black holes can radiate charge through the Schwinger effect as well.\footnote{It is particularly important to understand the possible decay channels of near-Nariai black holes through Schwinger pair production \cite{Montero:2019ekk,Aalsma:2023mkz}.} This motivates the utilization of an ensemble of de Sitter black holes that includes charge exchange to describe the thermodynamics of these black holes.

We already encountered two cases where this might be relevant: the intersection of the lukewarm and Nariai line and the ultracold point. In the first case, we found that the heat capacity at fixed charge diverges\footnote{As we will show in more detail later, the heat capacity at fixed charge at the lukewarm line is finite in general. The heat capacity only diverges at the intersection of the lukewarm and Nariai line.}. In the second case, variations away from the ultracold point require variations in both the charge and mass to remain within the sharkfin. 

\subsubsection{Heat Capacity at Fixed Charge-to-Mass Ratio}
As mentioned, the existence of the Schwinger effect necessitates the discussion of black hole thermodynamics in an ensemble where charge can vary. We will be interested in the heat capacity using Bousso-Hawking normalization, but with the constraint $\delta Q/\delta M = z $ when considering variations.\footnote{We could also have expressed this constraint in terms of $\delta Q/\delta\tilde M$ which would introduce an additional redshift factor.} Under this constraint, the heat flow through both horizons equals
\beq
\bal \label{eq:FirstLawZ}
\tilde T_b\delta S_b = +(1-z\Phi_b)\delta\tilde M ~, \\
\tilde T_c\delta S_c = -(1-z\Phi_c)\delta\tilde M ~.
\eal
\eeq
Note that the combination $z\Phi_{b,c}$ appearing here does not have a tilde. For neutral emissions ($z=0$) the heat flow through the black hole horizon always equals minus the heat flow through the cosmological horizon. Including charge, this is no longer true. We can for example consider the situation where there is positive (or vanishing) heat flow through both horizons, which occurs when
\beq \label{eq:EqualFlow}
\tilde T_b\delta S_b = \tilde T_c\delta S_c \quad \Rightarrow \quad z = \frac2{\Phi_b+\Phi_c} ~.
\eeq
In the Nariai limit $\Phi_b=\Phi_c$ and at this point both heat flows vanish when \eqref{eq:EqualFlow} is satisfied.

Now remember that at the intersection of the lukewarm and Nariai line, the heat capacity at fixed charge diverged. This implies that, at that point, the heat flow through the black hole horizon is independent of the temperature. To obtain a finite value for the heat capacity, we need to allow for charge exchange such that the heat flow vanishes. Concretely, we can only obtain a finite value for the heat capacity when $z = \Phi_b^{-1}$, at which point we find
\beq
r_b = r_c = \frac{\ell_4}{2}:\quad \tilde C_b^z = \tilde T_b\left(\frac{\partial S_b}{\partial\tilde T_b}\right)_{z=\Phi_b^{-1}} = S_b ~.
\eeq
We now also return to the ultracold point. Similar to the case at the intersection of the lukewarm and Nariai line, the value of $z$ for which the right-hand side of \eqref{eq:FirstLawZ} vanishes is given by $z = \Phi_b^{-1}$. In the case of the ultracold black hole this translates to
\beq
r_b = r_c = \frac{\ell_4}{\sqrt{6}}: \quad z = 2\sqrt{2\pi G_4} ~,
\eeq
which is precisely the value corresponding to the purple trajectory described in Fig. \ref{fig:UltraLinePlot} necessary to consider variations that remain within the sharkfin. We summarize the heat capacities at fixed $z$ in Tab. \ref{tab:HeatCapPot}. 
\begin{table}[t]
\centering
\makegapedcells
\begin{tabular}{| r || c c ||} 
 \hline 
 & $C_b^z$ & $\tilde C_b^z$ \\
 \hline\hline
Ultracold: & 0  & 0  \\
 \hline
 Nariai $\cap$ Lukewarm: & 0  & $S_b$  \\
 \hline
\end{tabular}
\caption{Heat capacity at fixed charge-to-mass ratio $z=\Phi_b^{-1}$ for both $\gamma=1$ and Bousso-Hawking normalization. $S_b$ denotes the entropy of the black hole.}
\label{tab:HeatCapPot}
\end{table}

\subsection{A Detailed Look at the Lukewarm Line}
Having understood the different heat capacities relevant for different parts of the phase diagram, we will now take a closer look at lukewarm black holes. As we already noted, this is a regime of interest for various reasons.

Foremost, it is a thermal equilibrium state where the temperatures of the cosmological and black hole horizon are equal. A corollary of this fact is that lukewarm black holes admit regular Euclidean continuations. Furthermore, because lukewarm black holes mark the boundary between parts of the phase diagram where their temperature is larger or smaller than the temperature of the cosmological horizon they play an important role in determining the dynamic evolution in the sharkfin. Specifically, for fixed charge the lukewarm solution corresponds to a maximum (total) entropy state. This means that the semi-classical thermodynamical evolution will never be able to decrease the mass below the lukewarm value to reach the cold limit. This is because below the lukewarm point (at fixed charge), the cosmological horizon will have a higher temperature than the black hole and there is net energy flux into the black hole through Hawking radiation, increasing the black hole mass. In the past, this semi-classical evolution has been studied using Gibbons-Hawking normalization \cite{Montero:2019ekk,Bhattacharjee:2025wfv}. Here we will study the behavior of black hole states near the lukewarm line using Bousso-Hawking normalization instead.

Lukewarm black holes are defined by
\beq \bc
f(r_b)=f(r_a)=0\\
f'(r_b)=-f'(r_c)
\ec ,\eeq
where the first equation defines the horizons and the second relates the temperature as
\beq
\tilde T_{b,c}=\frac{|f'(r_{b,c})|}{4\pi\sqrt{f(r_{\cal O})}} ~.
\eeq
The metric function thus has the following decomposition
\beq
f(r)=-\frac{1}{(r_b+r_c)^2r^2}(r-r_b)(r-r_c)\left[r^2+(r_b+r_c)r-r_b r_c\right] ~,
\eeq
from which one can read off the parameters
\beq
\bal
\ell_4&=r_b+r_c ~,\\
\mc{M}&=\frac{r_b r_c}{\ell_4}~,\\
\mc{Q}^2&=\frac{r_b^2 r_c^2}{\ell_4^2} ~,
\eal
\eeq
i.e., lukewarm black holes satisfy the condition $\mc{M}=|\mc{Q}|$. 

Because the Hawking temperature of the black hole and the cosmological horizon is the same, one would expect the lukewarm black holes to be in equilibrium when the spacetime is filled with Hawking radiation. This equilibrium will be broken
by the emission of charged particles through the Schwinger effect, subsequently putting the system in an ensemble satisfying $\delta\mc{Q}/\delta\mc{M} =\mc{Z}$, where the $\mc{Z}$ is determined by the charged particle in question.\footnote{The fixed charge-to-mass ratio constraint can be written in terms of $z=\delta Q/\delta M$ or $\mc{Z}=\delta{\cal Q}/\delta {\cal M}$. We'll use them interchangeably.} 

In fact, the Schwinger effect always occurs for lukewarm black holes for any charged particle spectrum, unlike cold black holes where there exists a threshold value for the charge-to-mass ratio. The presence of a threshold charge-to-mass ratio has been used as a motivation for the Weak Gravity Conjecture \cite{Lin:2024jug,Lin:2025wfe}. The absence of a threshold value can be seen from the near-horizon structure of the lukewarm black hole. 

We expand the metric function near the black hole as
\beq
f(r)=C\rho +{\cal O}(\rho^2),
\eeq
where $\rho=r-r_b$ and $C=f'(r_b)=\frac{2(r_c-r_b)}{(r_b+r_c)^2}$. The metric to leading order is
\beq \bal
ds^2&=-C\rho\, \rmd t^2+\frac{\rmd\rho^2}{C\rho}+(r_b+\rho)^2\rmd\Omega_2^2 \\
&\simeq -\alpha^2x^2 \rmd t^2+\rmd x^2+r_b^2 \rmd \Omega_2^2 ~,
\eal \eeq
where we made the change of coordinates $\rho=\frac{C}{4}x^2$ and defined the acceleration $\alpha=\frac{C}{2}$. The above metric describes the local ${\rm Rindler}_2\times S^2$ space. Similarly, the near-horizon geometry of the cosmological horizon can also be identified as ${\rm Rindler}\times S^2$, but with a different $S^2$ size. 

Using the near-horizon expansion, the gauge field near the black hole horizon can be expressed as 
\beq \bal
A_t=-\frac{Q}{4\pi r} &\simeq A_t^{(0)}+\frac{Q}{4\pi r_b}\frac{\rho}{r_b}\\
&=A_t^{(0)}+\frac{Q}{4\pi}\frac{C}{4 r_b^2}x^2,
\eal \eeq
where $A_t^{(0)}$ is a constant piece. Since $F=\rmd A$ is proportional to the two-dimensional volume form $\sqrt{-g_2}\rmd^2x$, there is a constant electric field $E=\frac{Q}{4\pi r_b^2}$ in the near-horizon two-dimensional Rindler space. The same holds true at the cosmological horizon, just with a smaller electric field strength.

The Schwinger production rate is non-zero in the presence of charged fields, as can be inferred from the isometry between Rindler space and Minkowski space. In fact, the instantons that give rise to the Schwinger rate are isometric to those in Minkowski space. This is true because the solutions to the equations of motion are preserved under coordinate transforms. The `classical' part of the Schwinger production rate is invariant, thus only the coefficient related to the quantum fluctuations, sensitive to the choice of vacua, is modified. The total production rate therefore takes the form of the Schwinger rate in Minkowski space, rescaled to take into account the acceleration of the Rindler observer.

Notably, the above implies that, unlike extremal black holes whose Schwinger effect feature a threshold value in the charge-to-mass ratio of the produced particle, lukewarm black holes emit charged radiation regardless of the particle spectrum. Depending on the charge-to-mass ratio of the particle, the Schwinger effect can either take the black hole above or below the lukewarm line, where neutral Hawking radiation can then drive the system back to the lukewarm solution.

\begin{figure}[t]
    \centering
    \includegraphics[width=0.7\linewidth]{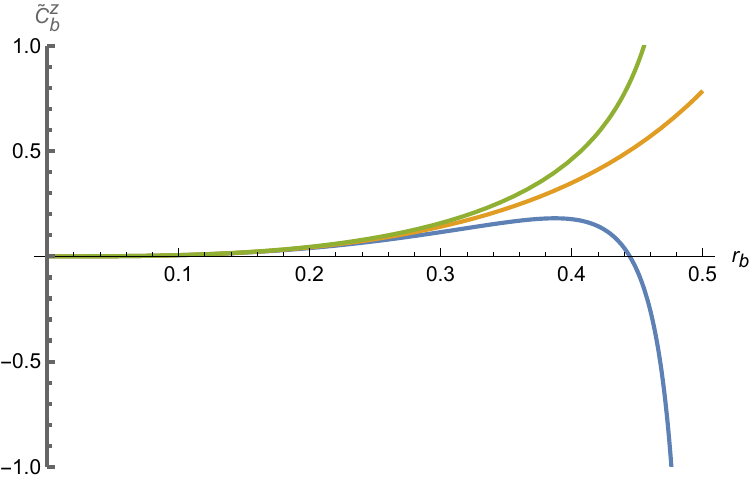}
    \caption{Heat capacity of lukewarm black holes in various ensembles specified by $\mc{Z}$. From top to bottom, the three curves correspond to $\mc{Z}=2.2,2,1.8$ respectively. When $\mc{Z}>2$, $\tilde{C}^z_b\rightarrow \infty$ at the lukewarm-Nariai intersection point whereas for $\mc{Z}<2$, $\tilde{C}^z_b\rightarrow -\infty$. For the latter case, $\tilde{C}^z_b$ has a zero at the point $1-\frac{\mc{Z} r_c}{\ell_4}=0$ which corresponds to $\delta S_b =0$.}
    \label{fig:lukewarm_Cb}
\end{figure}

Returning to the thermodynamics of lukewarm black holes, we will compute the heat capacity subject to the Schwinger effect. Using the general expression of the mass and charge parameters in terms of the black hole and cosmological horizons, the general variations are 
\beq \bc
\delta \mc{M} = \frac{r_b^2(r_b^2-\ell_4^2)-r_c^2(r_c^2-\ell_4^2)+2(2r_b^2-\ell_4^2)r_b(r_c-r_b)}{2\ell_4^2(r_c-r_b)^2}\delta r_b
-\frac{r_b^2(r_b^2-\ell_4^2)-r_c^2(r_c^2-\ell_4^2)+2(2r_c^2-\ell_4^2)r_c(r_c-r_b)}{2\ell_4^2(r_c-r_b)^2}\delta r_c\\
\delta (\mc{Q}^2)=\frac{1}{(r_c^{-1}-r_b^{-1})^2\ell_4^2}\left\{\left[(-3r_b^2+\ell_4^2)(r_c^{-1}-r_b^{-1})-\frac{r_c(r_c^2-\ell_4^2)-r_b(r_b^2-\ell_4^2)}{r_b^2}\right]\delta r_b \right.\\
\left. \qquad\qquad-\left[(-3r_c^2+\ell_4^2)(r_c^{-1}-r_b^{-1})-\frac{r_c(r_c^2-\ell_4^2)-r_b(r_b^2-\ell_4^2)}{r_c^2}\right]\delta r_c\right\}
\ec. \eeq
Restricting to the lukewarm line we obtain
\beq \bc
\delta \mc{M}=-\frac{1}{\ell_4^2}(r_b^2 \delta r_b + r_c^2\delta r_c) ~,\\
\delta \mc{Q} = -\frac{1}{\ell_4}(r_b \delta r_b+r_c \delta r_c) ~.
\ec \eeq
With the above expression, we can rewrite $\delta\mc{Q}= \mc{Z} \delta\mc{M}$ as
\beq \label{eq:var_relation}
\delta r_c= h(r_b,r_c)\delta r_b\equiv-\frac{r_b}{r_c}\frac{1-\mc{Z}\frac{r_b}{\ell_4}}{1-\mc{Z}\frac{r_c}{\ell_4}}\delta r_b.
\eeq
We next compute the heat capacity according to
\beq
\tilde{C}_b^z=\tilde{T}_b\left.\frac{\delta S_b}{\delta \tilde{T}_b}\right|_z ~,
\eeq
where the charge-to-mass ratio specifies the ensemble. We note that $S_b=\pi r_b^2/G_4$ and 
\beq
\tilde{T}_b=\frac{f'(r_b;r_b,r_c)}{4\pi \sqrt{f(r_{\cal O};r_b,r_c)}} ~.
\eeq
The semi-colon separates the variable $r$, from the parameters $r_{b,c}$. Later on, the parameter dependence will be left implicit. Derivatives of the radial coordinate will be denoted as a prime whereas derivatives with respect to the parameters will be denoted as partial derivatives. Further, when evaluating the metric function at a particular location, we will omit the argument and indicate this with a subscript. For instance, $f'_{\cal O}\equiv \frac{d}{dr}\left.f(r;r_b,r_c)\right|_{r=r_{\cal O}}$ is a function of $r_{b,c}$. Inserting these definitions we obtain
\beq
\tilde{C}_b^z=\frac{2\pi r_b f'_b}{\sqrt{f_{\cal O}}}\left[\frac{\delta}{\delta r_b}\left(\frac{f'_b}{\sqrt{f_{\cal O}}}\right)\right]^{-1} ~.
\eeq
While the quantities evaluated at $r=r_b$ can be simplified after restricting to the lukewarm line, the heat capacity in Bousso-Hawking normalization generally requires numerical evaluation due to the observer location $r_{\cal O}$ being the root of a quartic equation, which in general is a complicated lengthy expression. The behavior of $\tilde{C}^z_b$ with different choices of charge-to-mass ratio is shown in Fig. \ref{fig:lukewarm_Cb}.

From \eqref{eq:FirstLawZ}, we saw that $\delta S_b = 0$ when $z = \Phi_b^{-1}$ along the lukewarm line, which corresponds to $z = 2\sqrt{\pi G_4}\ell_4r_c^{-1}$. As a consequence the heat capacity also vanishes. However, this result does not directly indicate a breakdown of thermodynamics since we only allow decay channels that change the black hole charge-to-mass ratio in a constrained manner. In the canonical ensemble (fixed charge), a vanishing heat capacity reflects that the change in mass becomes independent of the temperature. Schwinger pair production on the other hand, does not depend directly on temperature. A vanishing heat capacity at fixed $z$ therefore does not reflect a similar breakdown of thermodynamics.

This different behavior is reflected in Fig. \ref{fig:lukewarm_Cb} where we can tune the charge-to-mass ratio to result in a zero in the heat capacity. While the heat capacity at fixed charge is always positive along the lukewarm line (away from the Nariai limit), allowing for charge exchange we see that the heat capacity at fixed charge-to-mass ratio can have both signs.

One intriguing aspect of working at fixed $z$ is that it allows one to tune the approach to the intersection of the lukewarm and Nariai line. As already mentioned, working at fixed charge this point has a diverging heat capacity suggestive of a continuous phase transition. To see this, we set ${\cal Z}=1$ to define an ensemble of lukewarm black holes. This allows us to approach the lukewarm-Nariai intersection while staying on the lukewarm line and track the change in energy and temperature by evaluating the heat capacity. For continuous phase transitions, the heat capacity has a power law divergence, the power of which defines a critical exponent that can be used to distinguish universality classes. The temperature of lukewarm black holes near the intersection is expanded as
\beq
\tilde{T}_b(r_b)=\tilde{T}_\text{crit}\left[1-\left(\frac{r_b}{\ell_4}-\frac{1}{2}\right)^2+\cdots\right],
\eeq
and the entropy is
\beq
S_b(r_b)=\frac{\pi \ell_4^2}{4G_4}+\frac{\pi \ell_4^2}{G_4} \left(\frac{r_b}{\ell_4}-\frac{1}{2}\right)+\cdots.
\eeq
From these two expansions, we see that the heat capacity diverges as
\beq
C_\text{LW}(r_b)=\tilde{T}_b\left.\frac{\delta S_b}{\delta \tilde{T}_b}\right|_\text{LW}\propto \left(\frac{r_b}{\ell_4}-\frac{1}{2}\right)^{-1} ~,
\eeq
such that we find
\beq
C_\text{LW}\propto\left(\frac{\tilde{T}_\text{crit}-\tilde{T}_b}{\tilde{T}_\text{crit}}\right)^{-\frac{1}{2}},
\eeq
indicating a critical exponent of $\alpha=\frac{1}{2}$. This exponent is associated to the divergence of the macroscopic response function $\tilde{C}_b$ and provides part of the information about the universality class of the lukewarm-Nariai critical point. The full information requires also microscopics of the black holes near criticality, for instance, the order parameter, correlation function and correlation length of gravitational fluctuations.

\section{Near-Extremal Expansion} \label{sec:NearExt}
Before we conclude, in this section we will give a complete comparison of the different thermodynamic quantities using the $\gamma=1$ and Bousso-Hawking normalization. We will focus on the different near-extremal limits ((charged) Nariai, cold) in an ensemble of fixed charge and tabulate the behavior of various thermodynamic quantities in Table \ref{tab:AllThermo}. The ultracold case will be treated separately.

The general approach we adopt is to expand the different horizon radii and $r_{\cal O}$ around their near-extremal values. 
\begin{equation}
	r_{a,b,c,\mathcal{O}}
	=
	\sum_{i=0}
	r_{ai,bi,ci,\mathcal{O}i}\epsilon^{i}
	\, ~,
\end{equation}
where $0\leq \epsilon \ll 1$ is a dimensionless and positive expansion parameter. This approach is complementary to our previous results and will therefore also serve as a consistency check.

A subset of the coefficients in this expansion will be fixed by requiring variations of $\ell_4$ and the charge $Q$ to vanish, in addition to imposing $f'(r_{\cal O})=0$ up to the relevant power in $\epsilon$. We summarize the relevant findings per case below.

\paragraph{Uncharged Nariai Black Hole.}
Here we take $r_{\mathcal{O}0}=r_{b0}=r_{c0}=r_{0}$ and $r_{ai}=0$ for all $i\geq0$. 
Explicitly this ansatz becomes
\begin{equation}
	r_{a}=0
	\,,
	\quad
	r_{b}
	=
	r_{0}
	-
	r_0\epsilon
	+
	r_{b2}\epsilon^{2}
	+
	r_{b3}\epsilon^{3}
	\,,
\end{equation}
\begin{equation}
	r_{c}
	=
	r_{0}
	+
	r_{c1}\epsilon
	+
	r_{c2}\epsilon^{2}
	+
	r_{c3}\epsilon^{3}	
	\,,
	\quad
	r_{\mathcal{O}}
	=
	r_{0}
	+
	r_{\mathcal{O}1}\epsilon
	+
	r_{\mathcal{O}2}\epsilon^{2}
	+
	r_{\mathcal{O}3}\epsilon^{3}
	\,,
\end{equation}
where, without loss of generality, we used the rescaling freedom of $\epsilon$ and the other coefficients to put $r_{b1}=-r_0$. This ensures that $\epsilon \geq 0$ and dimensionless, since the black hole radius decreases when moving away from the Nariai limit. We neglect terms of $\mathcal{O}(\epsilon^{4})$.

We use the constraints $f'(r_{\mathcal{O}})=0+\mathcal{O}(\epsilon^{4})$ and $\delta \ell_{4}=\mathcal{O}(\epsilon^{4})$. We choose these to fix the coefficients of $r_{c}$ and $r_\mathcal{O}$, which become
\begin{equation}
	r_{c1}
	=
	r_0
	\,,
	\quad
	r_{c2}
	=
	-
	\frac13(r_0+3r_{b2})
	\,,
	\quad
	r_{c3}
	=
	\frac19(r_0+6r_{b2}-9r_{b3})
	\,,
\end{equation}
\begin{equation}
	r_{\mathcal{O}1}
	=
	0
	\,,
	\quad
	r_{\mathcal{O}2}
	=
	-
	\frac{r_0}{2}
	\,,
	\quad
	r_{\mathcal{O}3}
	=
	\frac16(r_0+6r_{b2})
	\,.
\end{equation}
 The thermodynamic quantities coming from the expansions above are presented in Table \ref{tab:AllThermo}, in agreement with our previous results. We emphasize that $\epsilon$ can be expressed in terms of any (dimensionless) near-extremal thermodynamic quantity.

\paragraph{Charged Nariai Black Holes.}
Akin to the uncharged Nariai case, we take $r_{\mathcal{O}0}=r_{b0}=r_{c0}=r_{0}$. The fact that we allow for charge translates to allowing for a non-vanishing value for $r_{a}$. 
Explicitly, this ansatz becomes
\begin{equation}
	r_{a}
	=
	r_{a0}
	+
	r_{a1}\epsilon
	+
	r_{a2}\epsilon^{2}
	+
	r_{a3}\epsilon^{3}
	\,,
	\quad
	r_{b}
	=
	r_{0}
	-
	r_0\epsilon
	+
	r_{b2}\epsilon^{2}
	+
	r_{b3}\epsilon^{3}
	\,,
\end{equation}
\begin{equation}
	r_{c}
	=
	r_{0}
	+
	r_{c1}\epsilon
	+
	r_{c2}\epsilon^{2}
	+
	r_{c3}\epsilon^{3}	
	\,,
	\quad
	r_{\mathcal{O}}
	=
	r_{0}
	+
	r_{\mathcal{O}1}\epsilon
	+
	r_{\mathcal{O}2}\epsilon^{2}
	+
	r_{\mathcal{O}3}\epsilon^{3}
	\,,
\end{equation}
where again rescaled $r_{b1}$ without any loss of generality and consistent with $\epsilon\geq 0$ and dimensionless. On top of requiring $f'(r_{\mathcal{O}})=0+\mathcal{O}(\epsilon^{4})$ and $\delta\ell_{4}=\mathcal{O}(\epsilon^{4})$ to fix the coefficients of $r_{\mathcal{O}}$ and $r_{c}$, we impose $\delta Q=\mathcal{O}(\epsilon^{4})$ in order to fix the coefficients of $r_{a}$ and enforce the canonical ensemble. The explicit expressions for these coefficients are straightforward to obtain, but complicated and not very insightful. We will therefore not present them here\footnote{We are happy to provide them upon request.}. After using the expansions above as input we obtain the thermodynamic quantities as presented in Table \ref{tab:AllThermo}.

Note that $3 r_{0}^2\leq \ell^{2}_{4}\leq 6r_{0}^{2}$, where the lower (upper) limit corresponds to the uncharged Nariai (ultracold) limit. The value $4r_{0}^{2}=\ell^{2}_{4}$ corresponds to the intersection of the lukewarm and Nariai line at which point $\tilde C_b^Q$ diverges, which we already discussed before. The different heat capacities agree with our previous results.

\paragraph{Cold Black Holes.}
For cold black holes, at leading order in $\epsilon$ we find that $f(r_{\cal O})$ is constant, rather than $\epsilon^{-1}$. This means that $C^Q_b$ and $\tilde{C}^Q_b$ will be identical to leading order in $\epsilon$, both going to zero as $\epsilon\to0$. Thus, a possible breakdown of thermodynamics can also be determined using $C^Q_b$ which has been computed previously in \cite{Castro:2022cuo,Blacker:2025zca}.

The expansion takes the form
\begin{equation}
	r_{a}
	=
	r_{0}
	+
	r_{a1}\epsilon
	+
	r_{a2}\epsilon^{2}
	+
	r_{a3}\epsilon^{3}
	\,,
	\quad
	r_{b}
	=
	r_{0}
	+
	r_0\epsilon
	+
	r_{b2}\epsilon^{2}
	+
	r_{b3}\epsilon^{3}
	\,,
\end{equation}
\begin{equation}
	r_{c}
	=
	r_{c0}
	+
	r_{c1}\epsilon
	+
	r_{c2}\epsilon^{2}
	+
	r_{c3}\epsilon^{3}	
	\,,
	\quad
	r_{\mathcal{O}}
	=
	r_{\mathcal{O}0}
	+
	r_{\mathcal{O}1}\epsilon
	+
	r_{\mathcal{O}2}\epsilon^{2}
	+
	r_{\mathcal{O}3}\epsilon^{3}
	\,.
\end{equation}
Note that we now took $r_{b1} = r_0$ since the horizon radius of cold black holes increases when moving away from extremality. We note that $\ell_{4}^{2}\geq 6 r_{0}^{2}$.

Again, the coefficients are fixed by imposing $f'(r_{\mathcal{O}})=0+\mathcal{O}(\epsilon^{4})$,  $\delta\ell_{4}=\mathcal{O}(\epsilon^{4})$ and $\delta Q=\mathcal{O}(\epsilon^{4})$. As for the charged Nariai case, the explicit expressions are rather long and we omit them here. The location of $r_{{\cal O}0}$ can be determined via the equation
\beq
\ell_4^2=\frac{r_{{\cal O}0}^3}{r_0}+r_0 r_{{\cal O}0}+3 r_0^2+r_{{\cal O}0}^2 ~.
\eeq
The results are displayed in Table \ref{tab:AllThermo}, but we denote a few of the more lengthy quantities separately here.
\beq
\bal \label{eq:ColdValues}
\tilde T_b^{\rm cold} &=  \frac{\epsilon(\ell_4^2-6r_0^2)}{2\pi\ell_4^2r_0}\sqrt{\frac{r_{{\cal O}0}r_0\ell_4^2}{(r_{{\cal O}0}-r_0)^3(r_{{\cal O}0}+r_0)}} + {\cal O}(\epsilon^2) ~, \\
\tilde T_c^{\rm cold}
&= \frac{\left(\sqrt{\ell_4^2-2 r_0^2}-2 r_0\right) \left(\ell_4^2-2 r_0 \left(\sqrt{\ell_4^2-2 r_0^2}+r_0\right)\right)}{2 \pi  \ell_4 \left(r_0-\sqrt{\ell_4^2-2 r_0^2}\right){}^2}\sqrt{\frac{r_0 r_{{\cal O}0}}{(r_{{\cal O}0}-r_0)^3 (r_0+r_{{\cal O}0})}} + {\cal O}(\epsilon) ~, \\
\tilde \Phi_b^{\rm cold} &= \frac{\sqrt{r_0^2-\frac{3 r_0^4}{\ell_4^2}} \left(r_{{\cal O}0} (1-\epsilon )-r_0\right)}{2 \sqrt{\pi G_4 } r_0 r_{{\cal O}0}}\sqrt{\frac{\ell_4^2 r_0 r_{{\cal O}0}}{\left(r_{{\cal O}0}-r_0\right){}^3 \left(r_0+r_{{\cal O}0}\right)}} + {\cal O}(\epsilon^2) ~, \\
\tilde \Phi_c^{\rm cold} &= -\frac{r_0 \left(r_{{\cal O}0} \left(r_0+r_{{\cal O}0}\right)-\sqrt{r_0 \left(r_0+r_{{\cal O}0}\right) \left(r_0^2+r_{{\cal O}0}^2\right)}\right)}{2 \sqrt{\pi G_4\left(r_{{\cal O}0}-r_0\right){}^3 \left(r_0+r_{{\cal O}0}\right) \left(r_0 r_{{\cal O}0}+r_0^2+r_{{\cal O}0}^2\right) }} + {\cal O}(\epsilon)~.
\eal
\eeq

\paragraph{Ultracold Black Hole.}
The ultracold case is more subtle because, as explained, it is neccesary to consider trajectories that obey $\delta Q/\delta M = 2\sqrt{2\pi G_4}$. Instead of following the previous approach of fixing coefficients in a general fashion we note that an expansion of the form
\beq
r_b = \frac{\ell_4}{\sqrt{6}} - w_2\epsilon ~, \quad  r_c = \frac{\ell_4}{\sqrt{6}} +(w_1+w_2)\epsilon ~, \quad r_{\cal O} = \frac{\ell_4}{\sqrt{6}} +\sqrt{\frac{w_1^2+w_1w_1+w_2^2}{3}} ~.
\eeq
solves the constraints $f'(r_{\cal O}) = 0 + {\cal O}(\epsilon^3) $ and $\delta Q/\delta M = 2\sqrt{2\pi G_4} + {\cal O}(\epsilon^3)$.

Using these expansions, all relevant thermodynamic quantities can be obtained in a straightforward manner. Instead of giving a full list, we just highlight the most important results. Due the fact that $\sqrt{f(r_{\cal O})} = {\cal O}(\epsilon^{3/2})$ we find that the temperatures have a different scaling with $\epsilon$. For $\gamma = 1$ normalization we have $T_{b,c} \sim \epsilon^2$ whereas for Bousso-Hawking normalization $\tilde T_{b,c} \sim \epsilon^{1/2}$. Furthermore, the chemical potentials with Bousso-Hawking normalization diverge in the ultracold limit (the (physical) electric field remains finite). Lastly, we note that the near-extremal value of the heat capacity is equal in both normalizations and given by
\beq
C_b^z = \tilde C_b^{z} = -\sqrt{\frac23}\frac{\pi \ell_4 w_2\epsilon}{G_4} + {\cal O}(\epsilon^2) ~,
\eeq
where $z=2\sqrt{2\pi G_4}$.

\begin{table}[t]
\centering
\makegapedcells
\begin{tabular}{| c || c | c | c |} 
 \hline 
 & Uncharged Nariai & Charged Nariai & Cold \\ \hline \hline
 $S_b$ & $  \frac{\pi r_0^2}{G_4}\left(1-2\epsilon\right)$ & $  \frac{\pi r_0^2}{G_4}\left(1-2\epsilon\right)$ & $  \frac{\pi r_0^2}{G_4}\left(1+2\epsilon\right)$ \\
 $S_c$ & $  \frac{\pi r_0^2}{G_4}\left(1+2\epsilon\right)$ & $  \frac{\pi r_0^2}{G_4}\left(1+2\epsilon\right)$ & $  \frac{\pi  \left(\ell_4^2-r_0 \left(2 \sqrt{\ell_4^2-2 r_0^2}+r_0\right)\right)}{G_4} $ \\
 $T_b$ & $ \frac{\epsilon}{2\pi r_0}$ & $  \frac{\epsilon}{2\pi r_0}\left(\frac{6r_0^2}{\ell_4^2}-1\right)$ & $  \frac{\epsilon}{2 \pi  r_0}\left(1-\frac{6 r_0^2}{\ell_4^2}\right)$ \\
 $T_c$ & $ \frac{\epsilon}{2\pi r_0}$ & $  \frac{\epsilon}{2\pi r_0}\left(\frac{6r_0^2}{\ell_4^2}-1\right)$ & $  \frac{\sqrt{\ell_4^2-2 r_0^2} \left(\sqrt{\ell_4^2-2 r_0^2}-2 r_0\right)^2}{2 \pi  \ell_4^2 \left(r_0-\sqrt{\ell_4^2-2 r_0^2}\right)^2}$ \\
 $\tilde T_b$ & $ \frac{1}{2\pi r_0}\left(1+\frac23\epsilon\right) $ & $  \frac{\sqrt{6 r_0^2-\ell_4^2}}{2 \pi  \ell_4 r_0}\left(1+\frac{2 \epsilon  \left(\ell_4^2-4 r_0^2\right)}{\ell_4^2-6 r_0^2}\right)$ & $ \tilde T_b^{\rm cold} $  \\
 $\tilde T_c$ & $ \frac{1}{2\pi r_0}\left(1-\frac23\epsilon\right) $ & $  \frac{\sqrt{6 r_0^2-\ell_4^2}}{2 \pi  \ell_4 r_0}\left(1-\frac{2 \epsilon  \left(\ell_4^2-4 r_0^2\right)}{\ell_4^2-6 r_0^2}\right)$ & $\tilde T_c^{\rm cold}$ \\
 $\Phi_b$ & 0 & $  \sqrt{\frac{\ell_4^2-3 r_0^2}{4\pi G_4\ell_4^2}}(1+\epsilon )$ & $\sqrt{\frac{r_0\ell_4^2-3 r_0^3}{4\pi G_4r_0\ell_4^2}}(1-\epsilon)$ \\
 $\Phi_c$ & 0 & $  \sqrt{\frac{\ell_4^2-3 r_0^2}{4\pi G_4\ell_4^2}}(1-\epsilon )$ & $\frac{\sqrt{r_0^2\ell_4^2-3r_0^4}}{\sqrt{4\pi G_4\ell_4^2 } \left(\sqrt{\ell_4^2-2 r_0^2}-r_0\right)}$ \\
 $\tilde\Phi_b$ & 0 & $ \sqrt{\frac{\ell_4^2-3 r_0^2}{4\pi G_4(6r_0^2-\ell_4^2)}}\left(1-\frac{2 r_0^2 \epsilon }{\ell_4^2-6 r_0^2}\right)$  & $\tilde \Phi_b^{\rm cold}$  \\
 $\tilde\Phi_c$ & 0 & $ \sqrt{\frac{\ell_4^2-3 r_0^2}{4\pi G_4(6r_0^2-\ell_4^2)}}\left(-1-\frac{2 r_0^2 \epsilon }{\ell_4^2-6 r_0^2}\right)$ & $\tilde \Phi_c^{\rm cold}$  \\
 $Q^2$ & 0 & $ \frac{4 \pi  r_0^2 \left(\ell_4^2-3 r_0^2\right)}{G_4 \ell_4^2}$ & $\frac{4 \pi  r_0^2 \left(\ell_4^2-3 r_0^2\right)}{G_4 \ell_4^2}$ \\
 $\ell_4^2$ & $3r_0^2$ & $3r_0^2+2r_0r_{a0} + r_{a0}^2$ & $3r_0^2+2r_0r_{c0} + r_{c0}^2$  \\
 $C_b^Q$ & $  -\frac{2\pi r_0^2\epsilon}{G_4}$ & $  -\frac{2\pi r_0^2\epsilon}{G_4}$ & $  \frac{2\pi r_0^2\epsilon}{G_4}$ \\
 $\tilde C_b^Q$ & $  -\frac{3\pi r_0^2}{G_4}\left(1-\frac{11}6\epsilon\right)$& $ -\frac{\pi  r_0^2 \left(6 r_0^2-\ell_4^2\right)}{G_4 \left(4 r_0^2-\ell_4^2\right)}- \frac{\pi  r_0 \epsilon  \left(49 \ell_4^2 r_0^3-4 \ell_4^4 r_0-122 r_0^5\right)}{2 G_4 \left(\ell_4^2-4 r_0^2\right)^2}$  & $ \frac{2\pi r_0^2\epsilon}{G_4}$ \\
 \hline
\end{tabular}
\caption{Near-extremal expansion of the different thermodynamic quantities for (un)charged Nariai and cold black holes where we ignored terms of ${\cal O}(\epsilon^2)$ and smaller. Some of the more lengthy expressions are presented in \eqref{eq:ColdValues}. The results for the heat capacity agree with our previous results. The expansion parameter $\epsilon$ is dimensionless and positive. For the electric potential we took the gauge $\lambda = 0$ for $\gamma=1$ and $\tilde\lambda = r_{\cal O}^{-1}$ for Bousso-Hawking normalization.}
\label{tab:AllThermo}
\end{table}

\paragraph{Breakdown of the Semi-Classical Approximation.} 
From the above results, we can deduce where the thermodynamic description breaks down due to the (absolute value of the) heat capacity becoming smaller than one. Whenever this happens, one might expect large (quantum) corrections to modify the behavior of the partition function, for instance through log-$T$ corrections.

For cold black holes, the normalization of the Killing vector does not influence the value of the heat capacity to leading order in $\epsilon$. The heat capacity vanishes, the thermodynamic description breaks down and we expect log-$T$ corrections to the partition function.

For Nariai black holes, we found that sufficiently far away from the ultracold point the (absolute value of the) heat capacity remains large when using Bousso-Hawking normalization. This signals a consistent thermodynamic description and therefore the absence of large (quantum) corrections. This should be contrasted with $\gamma=1$ normalization where the heat capacity is zero along the entire Nariai branch. It would clearly be of great interest to compute the one-loop determinant of near-Nariai black holes using Bousso-Hawking normalization to see if this intuition is correct and whether the choice for the normalization of the Killing vector is important. Existing results on one-loop determinants for de Sitter black holes include \cite{Cotler:2019nbi,Maldacena:2019cbz,Maulik:2025phe,Turiaci:2025xwi,Blacker:2025zca,Arnaudo:2025btb,Law:2025yec} and some of these references did report large quantum corrections in the Nariai limit (see e.g. \cite{Maulik:2025phe,Turiaci:2025xwi,Blacker:2025zca} for recent results). These results, however, seem to depend on a complexification of the geometry describing the Milne patch of the near-horizon de Sitter region, instead of the static patch that we focus on. Therefore, these one-loop results apply in a different regime. Nonetheless, it would be interesting to better understand the relation between these computations and our thermodynamic approach.

Lastly, for ultracold black holes the value of the heat capacity is, again, the same for both choices of normalization. As a consequence we do expect a breakdown of the thermodynamic decscription and, most likely, large corrections. Explicitly computing these corrections in the ultracold limit, however, has remained a challenge.

\section{Conclusions} \label{sec:Conc}
In this work, we investigated the thermodynamics of four-dimensional electrically charged de Sitter black holes, focusing on the ambiguities that arise from the choice of normalization of the timelike Killing vector. We demonstrated explicitly that the conventional Gibbons-Hawking normalization leads to a non-standard interpretation of the black hole temperature; it expresses the surface gravity at the horizon as an acceleration measured in the frame of an accelerating observer that is located behind a horizon.

To remedy this, we adopted the Bousso-Hawking normalization of the Killing vector. This choice expresses the different thermodynamic quantities with respect to the frame of the unique freely falling observer at a fixed radial location in between the (outer) black hole horizon and cosmological horizon. This provides a more physically coherent picture, and we derived the corresponding first laws of thermodynamics for both the black hole and cosmological horizon. This led to the introduction of a new mass parameter $\tilde M$ whose variation equals the variation of the mass parameter $M$ (defined in the normalization used by Gibbons and Hawking) multiplied by a redshift factor that adapts it to the frame of the stationary free-falling observer.

Our primary result is the calculation of the black hole heat capacity using the Bousso-Hawking normalization, with a focus on its behavior near the three extremal limits: cold, Nariai, and ultracold. As established in the literature and discussed in the main text, a heat capacity that approaches order unity at low temperatures typically signals the breakdown of the semi-classical thermodynamic description. For asymptotically flat black holes, this phenomenon has been understood to stem from quantum corrections to the Euclidean path integral's saddle-point approximation. While these corrections are negligible at high temperatures, they dominate at low temperatures yielding the characteristic logarithmic (log-$T$) corrections to the partition function originating from fluctuations around the saddle-point that become dominant at zero temperature.

In contrast to results for the heat capacity using the normalization by Gibbons and Hawking, we find that the heat capacity remains finite and large for Nariai black holes, sufficiently far away from the ultracold point. For cold and ultracold black holes, however, we find the same result for the heat capacity, which vanishes in these extremal limits. Our results therefore suggest that away from the ultracold point there is no breakdown of the thermodynamic description along the Nariai branch.

For this reason, we also do not expect to see large (quantum) corrections in this limit, and it would be very interesting to confirm this prediction by an explicit one-loop computation that takes into account the physical normalization of the Killing vector. Intriguingly, the results of \cite{Arnaudo:2025btb} on Kerr-de Sitter black holes seem to point into this direction. Using the DHS formula \cite{Denef:2009kn}, the authors of \cite{Arnaudo:2025btb} identified whether log-$T$ corrections are present in the different extremal limits from the quasi-normal mode spectrum. Similar to our results, they find log-$T$ corrections for the cold and ultracold branch, but an absence of log-$T$ corrections for Nariai black holes. Although their approach focuses on rotating black holes and does not seem to rely on Bousso-Hawking normalization, we find this connection intriguing and we believe it warrants further investigation\footnote{We believe that their approach effectively implements our choice of normalization through the boundary conditions of the quasinormal modes.}.

It is also instructive to compare our findings with those of \cite{Maulik:2025phe,Blacker:2025zca}, which reported log-$T$ corrections for both cold and charged Nariai black holes. These studies derived corrections to the partition function by analyzing the (near-) zero modes of the Euclidean gravitational path integral. Specifically, in the Nariai limit, these zero modes are related to what is sometimes referred to as the large diffeomorphisms of the near-horizon de Sitter region. Since the appropriate evaluation of these modes involves a complexification of the geometry that takes one outside of the static patch, this seems to implicitly rely on a choice of reference frame different from ours. We therefore expect that their approach, phrased in terms of the heat capacity, corresponds to a normalization of the Killing vector different from the Bousso-Hawking one.

In summary, to fully understand the relation between the different approaches that study one-loop corrections to near-extremal black holes in de Sitter space, it is crucial to appreciate and understand their dependence on the selected normalization for the temperature, i.e. the particular physical observer that is being considered. This perspective seems to be in line with other recent developments that stress the role of the observer in de Sitter space \cite{Chandrasekaran:2022cip,Chen:2024rpx,Kudler-Flam:2024psh,Maldacena:2024spf,Speranza:2025joj,Chen:2025jqm}.

\subsection*{Acknowledgements}
We would like to thank Manus Visser for useful discussions. JPvdS would like to thank Zhenbin Yang for an interesting and insightful discussion at the ICTP Workshop on Approaches to Quantum Gravity in de Sitter space. LA is supported by the President's Postdoctoral Fellowship Program at the University of Minnesota and the Heising-Simons Foundation under the “Observational Signatures of Quantum Gravity” collaboration grant 2021-2818. LA thanks the organizers and participants of the workshop ``Quantum Gravity in Duluth'' where part of this work was presented and acknowledges the University of Wisconsin-Madison for hospitality. PL and GS are supported in part by the DOE grant DE-SC0017647. The work of WS is supported by Starting Grant 2023-03373 from the Swedish Research Council.

\appendix

\section{Quartic Polynomials} \label{app:quartic}
Here we give the general solution for the roots of a quartic polynomial. Consider a quartic polynomial given by
\beq
P_4 = ar^4 + br^3 + cr^2 + dr + e ~.
\eeq
The four roots $P_4=0$ are given by \cite{wiki:Quartic_function}
\beq
\bal
r_{1,2} &= - \frac{b}{4a}-S\pm\frac12\sqrt{-4S^2-2p+\frac qS} ~, \\
r_{3,4} &= - \frac{b}{4a}+S\pm\frac12\sqrt{-4S^2-2p-\frac qS} ~,
\eal
\eeq
with
\beq
\bal
p &= \frac{8ac-3b^2}{8a^2} ~, \\
q &= \frac{b^3-4abc+8a^2d}{8a^3} ~,
\eal
\eeq
and
\beq
\bal
S &= \frac12\sqrt{-\frac23p+\frac1{3a}\left(R+\frac{\Delta_0}{R}\right)} ~, \\
R &= \left(\frac{\Delta_1+\sqrt{\Delta_1^2-4\Delta_0^3}}{2}\right)^{1/3} ~,
\eal
\eeq
where
\beq
\bal
\Delta_0 &= c^2 - 3bd +12 ae ~, \\
\Delta_1 &= 2c^3 - 9bcd + 27b^2e +27ad^2 -72ace ~.
\eal
\eeq

\section{Derivation of Thermodynamic Quantities}\label{app:thermoquantities}
In this appendix, we specify the geometric setup for computing various thermodynamic quantities. 

\subsection{General Expressions}
We consider a $d$-dimensional spacetime and define a spacelike co-dimension 1 hypersurface $\Sigma$ with timelike unit normal $n_\mu$. We also consider a co-dimension 1 timelike surface $B$ with spacelike unit normal $r_\mu$. From the induced metric on $B$, we can construct the extrinsic curvature $k^{\mu\nu}$ with trace $k$. We use $\sigma_{\mu\nu}$ to denote the induced metric on the intersection $\Sigma \cap B$.

Following \cite{Banihashemi:2022htw}, for a timelike Killing vector $\xi^\mu$, we define the local acceleration associated with this trajectory by
\beq
a_\mu = \frac{\nabla_\mu(\sqrt{-\xi^\nu\xi_\nu})}{\sqrt{-\xi^\rho\xi_\rho}} ~.
\eeq
Interpreting $B$ as a York boundary, we define $s^{\mu\nu}$ to be the spatial stress contribution to the Brown-York stress tensor \cite{Brown:1992br}
\beq
s^{\mu\nu} = \frac1{8\pi G_d}\left(-k^{\mu\nu} + (r_\rho a^\rho+k)\sigma^{\mu\nu}\right) ~.
\eeq
Using this definition, we can define the pressure associated to the York boundary as
\beq
P_B = \frac1{d-2}s^{\mu\nu}\sigma_{\mu\nu} ~.
\eeq
Another quantity that we need to introduce is the anti-symmetric Killing potential $\omega^{\mu\nu}$ that satisfies
\beq
\nabla_\nu \omega^{\nu\mu} = \xi^\mu ~.
\eeq
Given a Killing vector, the Killing potential is non unique and can be modified by a term $\omega^{\mu\nu}\to\omega^{\nu}+\lambda^{\mu\nu}$, where $\lambda^{\mu\nu}$ is an anti-symmetric tensor that satisfies $\nabla_\nu \lambda^{\nu\mu}=0$. From the Killing potential, one can define the thermodynamic volume between two different timelike surfaces $B_1$ and $B_2$. For example, for $B_b$ the black hole horizon and $B_\infty$ infinity the thermodynamic volume is given by
\beq \label{eq:ThermoVol}
\int_{\Sigma\cap B_\infty}\rmd A\,r_\mu n_\nu\left(\omega^{\mu\nu}-\omega^{\mu\nu}_0\right) - \int_{\Sigma\cap B_b}\rmd A\, r_\mu n_\nu \omega^{\mu\nu} ~,
\eeq
where we included the possibility of adding a counter term $\omega^{\mu\nu}$ to subtract of a divergence. $\rmd A$ is the area element on the intersection $\Sigma\cap B$.

\subsection{Charged de Sitter Black Holes}
Now let us evaluate these quantities explicitly for four-dimensional charged de Sitter black holes with Killing vector $\xi = \gamma\partial_t$. We find that the pressure at a location of constant $r$ is given by
\beq
P_r^{-1} = 8\pi G_4\ell_4^2 r^2\sqrt{-\frac{\left(r-r_b\right) \left(r-r_c\right) \left(r_b \left(r_c+r\right)+r_b^2+r_c \left(r_c+r\right)-\ell_4^2+r^2\right)}{\ell_4^2 \left(r_b r_c \left(r_b r_c+r_b^2+r_c^2-\ell_4^2\right)+r^2 \left(\ell_4^2-3 r^2\right)\right){}^2}}
\eeq
To derive the Killing potential, we take the ansatz that the only non-zero components are $\omega^{tr} = - \omega^{rt}$ and we assume the components only depend on the radial coordinate. We then find
\beq
\omega^{tr} = \gamma\left(-\frac r3+\alpha\frac{r_b^3}{r^2}\right) ~,
\eeq
with $\alpha$ a constant that captures the ambiguity in the Killing potential. To compute the thermodynamic volume between the black hole horizon and infinity, we need to specify a counter term due to a divergence. We specify $\omega^{\mu\nu}_0$ to be the Killing potential for empty de Sitter space, which suggests us to set $\alpha=0$ to remove the divergence at $r=0$. Doing so, we find that \eqref{eq:ThermoVol} evaluates to
\beq
V_b = \gamma\frac43\pi r_b^3 ~.
\eeq
If we compute the thermodynamic volume between two surfaces $B_1$ and $B_2$ at (finite) radius $r=r_1$ and $r=r_2$ we find that there is no need to regularize, $\alpha$ drops out and we are left with
\beq
V_{r_2}-V_{r_1} = \gamma\frac43\pi(r_2^3-r_1^3) ~.
\eeq

\section{Energy in Bousso-Hawking normalization} \label{app:BHenergy}
In the main text, we introduced (the variation of) a new energy parameter $\tilde M$. Its definition might seem ad-hoc, so in this Appendix we clarify its interpretation. First, we highlight that from the variational identity
\beq
\delta\tilde M = \delta E + P_{\cal O}\delta A_{\cal O} + \tilde\Phi_{\cal O}\delta Q ~,
\eeq
we see that at fixed charge and York boundary, the variation $\delta \tilde M$ coincides with the variation of the Brown-York energy.

Second, the variational identity can be integrated to obtain an expression for $\tilde M$. To this end, it is useful to write down an ansatz
\beq
\tilde M = \frac{M}{\sqrt{f(r_{\cal O})}} + {\cal M} ~.
\eeq
Varying this ansatz, imposing \eqref{eq:NewMassVariation}, and integrating leads to
\beq
{\cal M} = \frac{M}{2f(r_{\cal O})^{3/2}}\left(\int\rmd M \,\partial_Mf(r_{\cal O})+\int\rmd Q\,\partial_Q f(r_{\cal O})\right) ~.
\eeq
Performing these integrals is complicated, but possible in certain specific cases. This gives an explicit expression for $\tilde M$.

It should be noted that it is not necessary to introduce $\tilde M$ to compute the heat capacity, as we can just use the definition in terms of the heat exchanged ($\tilde T\delta S$). We find the energy perspective useful, however, because it highlights that the heat capacity obtained using $\tilde M$ coincides with the heat capacity one would have obtained when using the Tolman temperature with a fixed York boundary and fixed charge. This is only true when the York boundary is located at $r=r_{\cal O}$. A consequence is that $\tilde M$, which we initially introduced in the main body of this paper, can be interpreted as the thermodynamic (Brown-York) energy that appears in the partition function in the presence of a York boundary.

This suggests that there are two different perspectives that give the same result for the thermodynamics. One can either absorb all dependence on the York boundary and allow for variations in $r_{\cal O}$ (which is the perspective we took), or impose a York boundary at $r=r_{\cal O}$ and fix its location. We'd like to stress that the equivalence of these two perspectives is \emph{only} true for our special choice of observer at $r=r_{\cal O}$. Mathematically, the reason is that a variation of $f(r_{\cal O})$ with respect to $r_{\cal O}$ vanishes both when we fix $r_{\cal O}$ (by definition) and also when we do allow for variations because $r_{\cal O}$ is defined to satisfy $f'(r_{\cal O}) = 0$. These observations support the claim that $\tilde M$ is the appropriate definition of energy with respect to an observer at $r_{\cal O}$.

\bibliographystyle{utphys}
\bibliography{refs_paper}

\end{document}